\documentclass{article}

\usepackage{arxiv}

\usepackage[utf8]{inputenc} 
\usepackage[T1]{fontenc}    
\usepackage{hyperref}       
\usepackage{url}            
\usepackage{booktabs}       
\usepackage{amsfonts}       
\usepackage{nicefrac}       
\usepackage{microtype}      
\usepackage{lipsum}		
\usepackage{graphicx}
\usepackage{natbib}
\usepackage{doi}

\usepackage{amssymb, bm}
\usepackage{amsmath}
\usepackage{tabularx}
\usepackage[titletoc]{appendix}
\usepackage{color}

\usepackage{comment}
\usepackage{newtxtext}
\usepackage{newtxmath}
\usepackage{lineno}
\usepackage{caption}
\usepackage{subcaption}

\title{A two-way coupled high resolution wave hindcast for the South China Sea}

\author{ \href{https://orcid.org/0000-0002-1390-2367}{\includegraphics[scale=0.06]{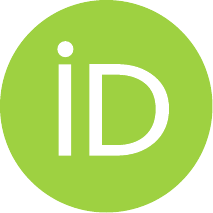}\hspace{1mm}Tiziano Bagnasco} \\
	Department of Civil and Environmental Engineering \\
	The Hong Kong Polytechnic University\\
	Hong Kong S.A.R., China \\
	\texttt{tiziano.bagnasco@connect.polyu.hk} \\
	\And
	\href{https://orcid.org/0000-0001-6087-4250}{\includegraphics[scale=0.06]{orcid.pdf}\hspace{1mm}Alessandro Stocchino}\thanks{corresponding author.} \\
	Department of Civil and Environmental Engineering \\
	The Hong Kong Polytechnic University\\
	Hong Kong S.A.R., China \\
	\texttt{alessandro.stocchino@polyu.edu.hk} \\
	\And
	\href{https://orcid.org/0000-0003-2655-6181}{\includegraphics[scale=0.06]{orcid.pdf}\hspace{1mm}Michalis I.Vousdoukas}\\
	University of the Aegean\\
	 Mytilene, Greece\\
	\texttt{vousdoukas@gmail.com} \\
	\And
	\href{https://orcid.org/0000-0003-1821-2362}{\includegraphics[scale=0.06]{orcid.pdf}\hspace{1mm}Jinghua WANG}\\
	Department of Civil and Environmental Engineering \\
	The Hong Kong Polytechnic University\\
	Hong Kong S.A.R., China \\
	\texttt{jinghua.wang@polyu.edu.hk} \\
}

\renewcommand{\shorttitle}{South China Sea wave hindcast}

\hypersetup{
pdftitle={A template for the arxiv style},
pdfsubject={q-bio.NC, q-bio.QM},
pdfauthor={David S.~Hippocampus, Elias D.~Striatum},
pdfkeywords={First keyword, Second keyword, More},
}

\begin{document}
\maketitle

\begin{abstract}
	In the present study, we performed a 53-year wave hindcast (1970-2022)  for a significant portion of the South China Sea (SCS) with an unstructured mesh that reaches considerably high resolution along the coasts of the Guangdong province (China). The adopted modeling approach is based on the fully two-way coupled SCHISM-WWMIII numerical suite. The model was forced with ERA5 wind velocities that were compared to IFREMER altimeter wind velocities and then bias-corrected for a more accurate treatment of the wind component. Eight major tidal harmonics extracted from FES2014 were imposed to the open boundaries. 
After a preliminary mesh independence analysis, the model results have been validated against satellite altimeter observations retrieved from the European Space Agency database spanning the period from 1992 to 2019. Moreover, 28 year in-situ measurements from two coastal wave buoys and data from four tidal gauge stations (approximately 20 years) were used to test the nearshore skills of the model. 
Several statistical indicators have been used to evaluate the offshore and nearshore performance of the model results in terms of the main wave parameters (significant wave height, peak wave period, mean wave direction) and water levels. All statistical metrics suggest that the present hindcast improved the predictions of waves and water levels compared to previous datasets, especially in the coastal regions. The high spatial resolution together with a full coupling allowed the model to capture and simulate processes that are induced by the non-linear interactions between waves and currents, especially nearshore. 
\end{abstract}


\section{Introduction}
\label{sec:Introdution}
An in-depth knowledge of the wave dynamics and the effects they induce are of primary importance in the effective design of nearshore/offshore structures, for the protection of coastal areas, for the wave energy assessment and for facilitating any marine operation. If the propagation of waves and their repercussions on beaches, coastal structures and piers can be anticipated, their adverse effects can be effectively mitigated \citep{di2018wave,saponieri2019evaluation}. Furthermore, given the limited availability of real-time wave measurements and observations, the 
implementation of numerical models to generate wave hindcasts can serve as a beneficial approach to comprehend the oceanic conditions and enhance wave forecasting for future scenario predictions.

In this regard, numerous numerical models can be utilized but their choice mainly depends on the purpose of the analysis. At present, the numerical models that deploy an unstructured mesh have become a valid alternative to models that use regular grids, especially for large-scale applications as stressed by \cite{mentaschi2023global}. This is because unstructured-grid models can generate detailed meshes that easily adapt to complex shorelines and sophisticated geometries. This becomes critical in areas characterized by numerous islands and profiles with irregular topography. The necessity of improving numerical simulations in such contexts, along with the need for higher-resolution results under constrasting spatial and temporal scales, has lead to the advent of unstructured-grid models \citep{dietrich2011modeling}.

Noteworthy examples of widely used circulation unstructured models include, for instance, the Finite-Volume Coastal Ocean Model (FVCOM, \cite{chen2003unstructured}), the System of HydrodYnamic Finite Element Modules (SHYFEM, \citep{federico2017coastal,Umgiesser2004}, ADCIRC \citep{luettich1992adcirc,lynch1996comprehensive}, TELEMAC model \citep{galland1991telemac} and the Semi-Implicit Cross-scale Hydroscience Integrated System Model (SCHISM,\citep{zhang2008selfe,zhang2016seamless,zhang2023global}).\ 
Concerning wave modeling, among the frequently employed spectral wave models are SWAN \citep{booij1999third}, TOMAWAC  \citep{benoit1997development}, WAVEWATCH III \citep{tolman1991third}, WAM \citep{group1988wam}, MIKE21 SW \citep{sorensen2005third},CREST \citep{ardhuin2001hybrid} and WWMIII \citep{roland2008development,roland2009development,roland2012fully}.

The existing literature provides several examples and applications of both global-scale wave hindcasts \citep{hemer2013global,perez2017gow2,stopa2019sea,mentaschi2023global}
and more regionally focused studies \citep{mirzaei2013wave,liang2016wave,shi201939}. The Eastern part of the North Atlantic coast was studied by \cite{pilar200844} and led to the generation of a 44-year wave hindcast (1958-2001) with the WAM model. 
\cite{mentaschi2013developing} analyzed the performance of WAVEWATCH III in the Western Mediterranean Sea concerning seventeen storm events. The Central and South Pacific was investigated by \cite{durrant2014global} while generating a 31-year global wave hindcast with the deployment of WAVEWATCH III. \cite{perez2017gow2} also generated a global wave hindcast from 1979 to 2015 with improved resolution of 0.25$^\circ$ in coastal areas and near the poles. 
Only later, \cite{mentaschi2023global} developed a 73-year global hindcast of waves and storm surges utilizing SCHISM-WWMV and considering an unstructured mesh characterized by more than 650000 nodes and with highest resolutions ranging between 2 and 4 kilometers near costal areas. 

The area examined in this paper encompasses a substantial portion of the SCS: a marginal sea of the Western Pacific Ocean enclosed by the Indochinese Peninsula, China, Taiwan, Philippines, Indonesia, Malaysia and Borneo. During the summer and autumn seasons, this region experiences the impact of intense winds and typhoons, occasionally leading to significant harm to coastal structures, offshore installations, and vessels \citep{shi2017development,wang2017improvements}. Extensive research efforts have been undertaken in this region to develop forecasting or hindcasting of waves, analyze wave climate trends, and achieve a comprehensive understanding of the wave dynamics. Several studies focused on wave condition analysis and wave energy assessment \citep{wang2018long,jiang2019assessment,sun2020assessment,yang2020long}. Researchers have also been exploring the effects of concurrent storm-tide-tsunami events \citep{wang2021numerical} and the observations and modelling of typhoon waves \citep{yang2015observed,xu2017observations,wu2018evaluation}. The literature also includes examples of wave hindcasts produced in this region.
\cite{mirzaei2013wave} conducted a 31-year wave hindcast using the WAVEWATCH III model, aiming to evaluate the long-term alterations and inter-annual variations in the wave climate of the SCS. The hindcast employed a minimum resolution of 0.3 degrees to depict the wave characteristics accurately. The wave climate in the Bohai Sea, Yellow Sea, and East China Sea for the period spanning 1990 to 2011 was investigated by \cite{liang2016wave} using the SWAN model and a minimum resolution of 1 arcminute. Another study by \cite{shi201939} produced a wave hindcast covering the time span from 1979 to 2017 along the Chinese coast reaching resolutions of 1km in shallow areas deploying the TOMAWAC model. 

Numerical wave modeling can be performed accounting for the influence and interaction of currents or by neglecting them entirely.
 As mentioned by \cite{kong2024model}, the exchange of data in two-way coupled models between waves and currents results in a more accurate representation of some real complex phenomena such as wave-induced turbulent mixing \citep{kumar2012implementation}. Practice has shown that, typically, coupled models exhibit superior performance in simulating wave conditions compared to wave-only models \citep{kong2024model}. This improvement can be attributed to the inclusion of current effects, which play a crucial role in accurately depicting phenomena such as dissipation, wave breaking, and wave steepening \citep{ardhuin2012numerical}. 
 
 For this study, we relied on the 2-way fully coupled SCHISM-WWMIII models, the specifics of which will be described in the next sections.
 The primary objective of this analysis is to produce a 53-year wave hindcast for a significant portion of the South China Sea, with a specific focus on the coastline of the Guangdong province (China), which stands as one of the most densely populated coastal areas. The numerical predictions in terms of wave characteristics and water levels will be validated against offshore satellite altimeter observations, coastal wave measurement stations and tidal gauges.  

\section{Material and Methods}\label{sec:Material and Methods}

\subsection{Numerical Model Set-up}\label{sec:Numerical Model Set-up}
\begin{figure}[h!]
\centerline{
\includegraphics[width=1\textwidth]{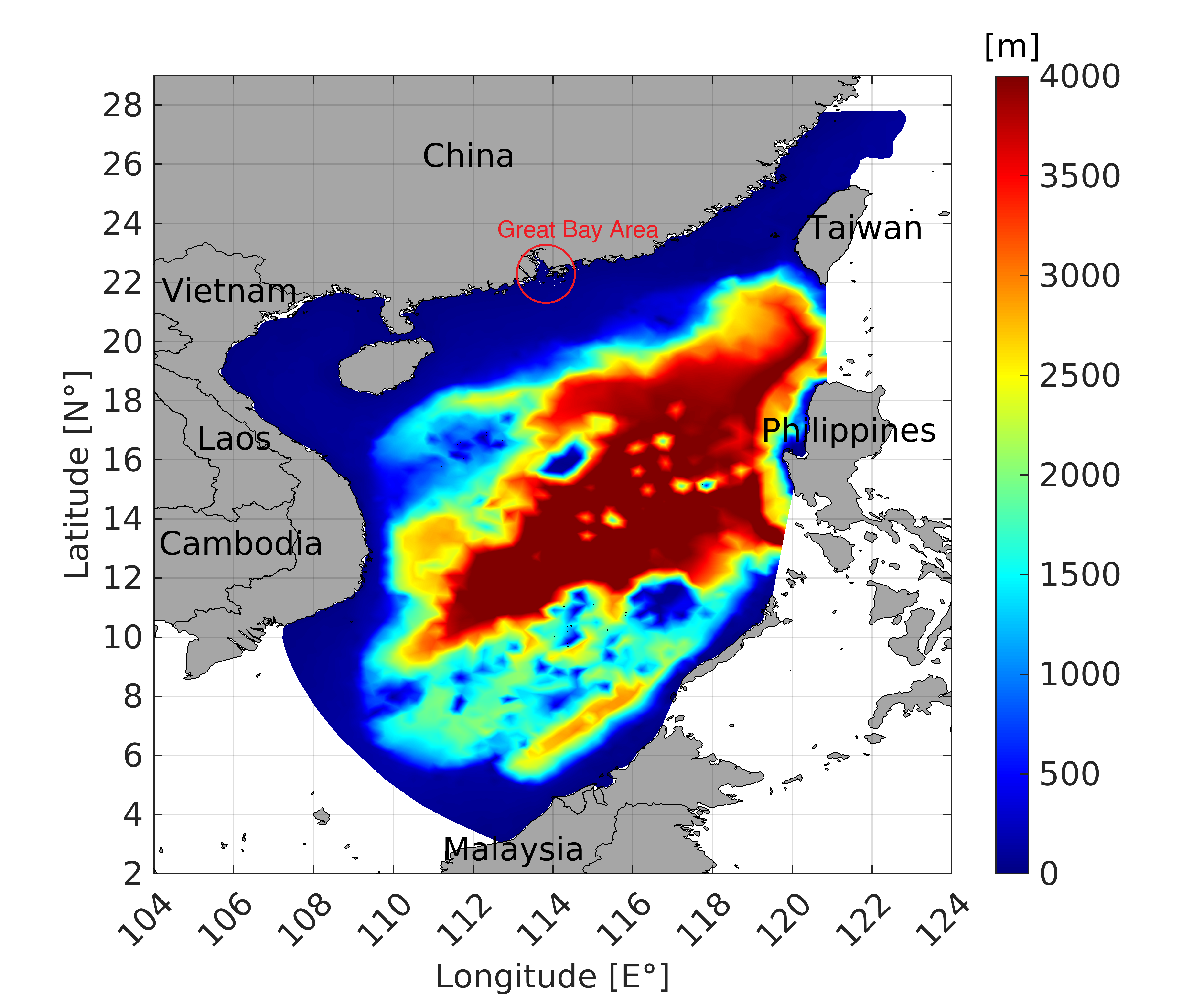}
}
\caption{Portion of the South China Sea investigated with the bathymetry contours.}
\label{fig:domain}
\end{figure}

The domain considered in this paper covers the majority of the SCS, spanning from Longitudes of approximately 106$^\circ$degree E to 123$^\circ$ E and from Latitudes of approximately 3$^\circ$ N to 28$^\circ$ N. Figure \ref{fig:domain} shows the part of the South China Sea considered for the numerical simulations. A particular focus was dedicated to the Great Bay Area (GBA) and the coasts along the Guangdong Province (China).

This study is based on a 2-dimensional, 2-way coupled model, combining the Semi-implicit Cross Scale Hydroscience Integrated System Model SCHISM and the third-generation spectral Wind Wave Model WWMIII.
The 2-way coupling implies that the current velocities and sea water levels are provided to the wave model and the latter computes the wind stresses that are then handed back to SCHISM.
The structure of SCHISM is derived from the original Semi-implicit Eulerian-Lagrangian Finite-Element model (SELFE), of which further details can be retrieved in  \cite{zhang2008selfe} and \cite{zhang2016seamless}. Since the interest of the study is to assess the interaction between wind waves and currents, no particular attention is dedicated along the vertical direction. For this purpose, SCHISM applies the 2D depth-integrated barotropic equations.
The integration time steps implemented in the analysis are 100s for SCHISM and 600s for WWMIII: this implies that the two models exchange information every 6 time steps. This choice was made after testing different model time steps in order to minimize possible numerical dispersion errors and/or excessive truncation errors. Being SCHISM a semi-implicit model itself that applies no mode splitting, the fulfillment of the CFL constraint is more easily met and this leads to greater numerical stability.  As stressed by \cite{huang2022tidal}, the typical integration time steps used in these types of field applications usually vary between 100s and 200s. The unstructured mesh covers the majority of the South China Sea (SCS) and was generated with the aid of the 13.1 version of SMS (Surface-water Modeling System by Aquaveo, https://www.xmswiki.com/wiki/SMS:SMS).
Both SCHISM and WWMIII share the same unstructured mesh which comprises 15523 nodes and 29039 triangular elements.
The lowest mesh resolution is circa 0.008$^\circ$ around the GBA and reaches a size of 0.35$^\circ$ in the open ocean.
Several unstructured meshes were generated and tested in order to find the optimal mesh size that can better capture the wave propagation nearshore in the Hong Kong waters. The bathymetry of the whole domain was extracted from the 2022 General Bathymetric Chart of the Oceans (GEBCO) dataset \citep{weatherall2015new} with a resolution of 15 arc-second intervals and then merged with a more refined dataset of water depths related to Hong Kong only.  

The bottom friction was evaluated considering a minimum boundary layer thickness of 0.2m and a roughness length equal to $5.78\cdot10\textsuperscript{-6}$m and constant for all the nodes. The latter corresponds to a Manning coefficient of about $0.012$ s/m\textsuperscript{1/3} which was also applied by \cite{wang2021numerical} for a similar application in the SCS. The atmospheric forcing (sea level pressure and wind velocities) and the tidal oscillations at the open boundaries start from a zero value and then have a warm-up time equal to 365 days in order to provide enough time for the Oceanic conditions to settle properly in such a large domain. The wetting and drying mechanism was considered active and accounted for a minimum water depth of 0.01m. The surface stress was calculated according to the parameters of \cite{donelan1993dependence}, whose formulation is based on the wave age. The discretization of the spectral domain was performed regrouping the frequencies into 36 bins ( lower limit of 0.04 Hz and higher limit of 1 Hz) and the directions into 24 bins with minimum and maximum directions of respectively 0$^\circ$ and 360 $^\circ$. The wave boundary layer was activated and treated as explained by \cite{soulsby1997dynamics}. WWMIII solves the wave action equation and accounts for different contributions and phenomena, mainly referred to as source terms, that cause the energy content to be rearranged within the spectrum. The work done in this paper considers non-linear interactions (DIA approximation,\citep{hasselmann1985computations}), wind-induced energy input and wave energy dissipation in deep waters (whitecapping) according to the ST4 parametrization \citep{ardhuin2010semiempirical}. The model incorporates the JONSWAP bottom friction parametrization \citep{hasselmann1973measurements} with a bottom friction coefficient equal to 0.067 m\textsuperscript{2}s\textsuperscript{-3} and shallow water wave breaking characterized by a constant gamma criterion. The wave breaking formulation is based on the work of \cite{battjes1978energy} and the model also incorporates the effects of triad 3-wave interactions (LTA Lumped Triad Approximation,\citep{eldeberky1996nonlinear}).

The numerical simulations aimed at generating a 53-year wave hindcast cover the time period between the years 1970 and 2022. The outputs of both SCHISM and WWMIII are generated and saved hourly and comprise the main variables used for the purpose of this paper: significant wave height ($H_s$), peak wave period ($T_p$), mean wave direction ($D_m$) and water level ($h$).

\subsection{Meteocean forcing and open boundary conditions}\label{sec:Study area and meteocean forcing data}
\begin{figure}[h!]
\centerline{
\includegraphics[width=1\textwidth]{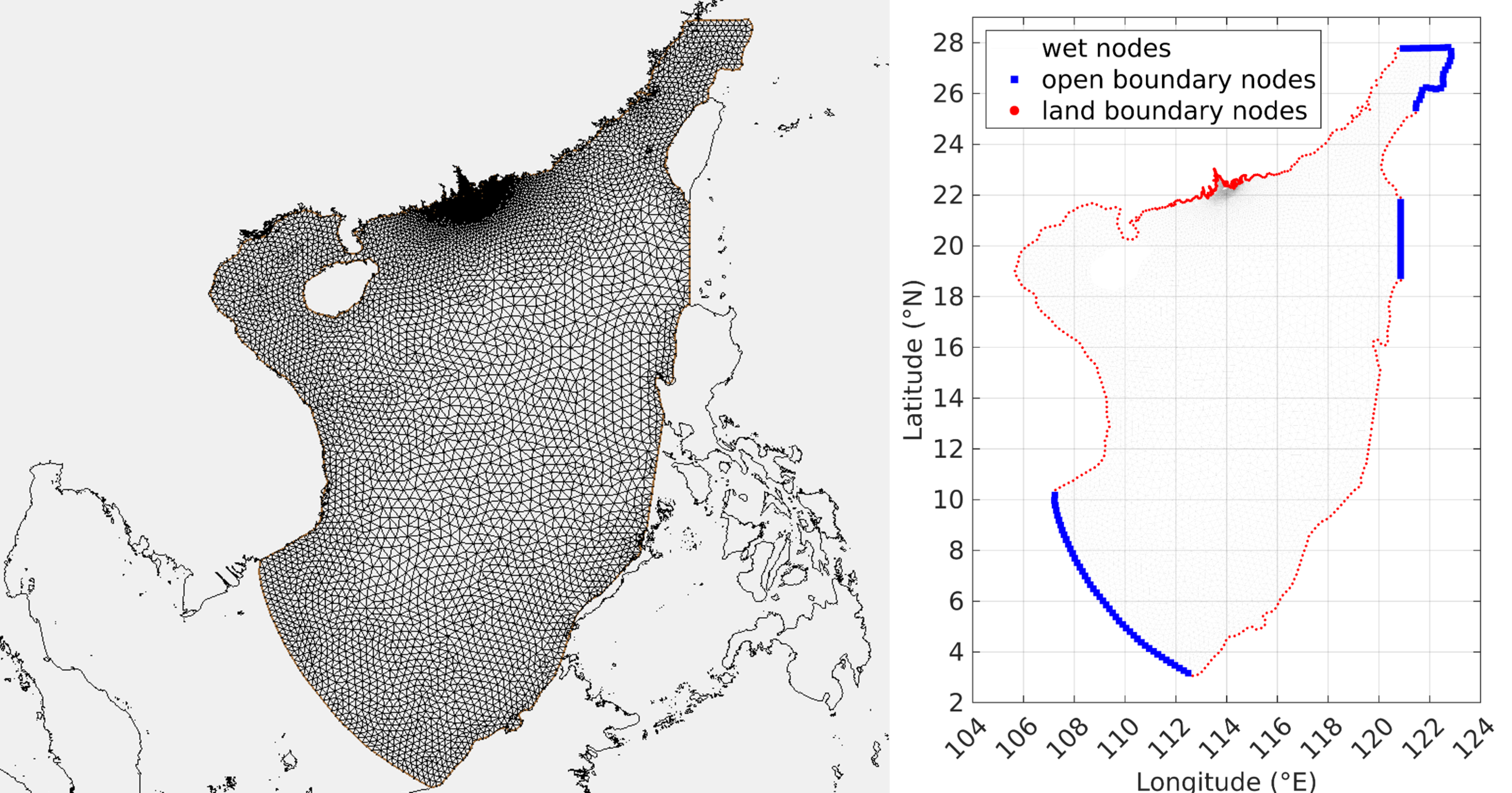}
}
\caption{Left panel: mesh used for the numerical simulations. Right panel: Details of the unstructured mesh and boundaries of the numerical domain}
\label{fig:mesh}
\end{figure}

The boundary conditions applied at the open boundary nodes of the unstructured mesh (blue nodes in Figure \ref{fig:mesh} right panel) consist of the 8 major harmonic components of the tides ( M2, S2, N2, K2, K1, O1, Q1, P1) and were extracted from the FES2014 package {\citep{carrere2015fes}}. The boundary nodes that connect Malaysia to the region above Manila in the Philippines were considered as land nodes since the portion of water that they exclude is enclosed by various islands and therefore not interesting for wind waves generation. The main meteorological forcing at stake are hourly data of sea level pressure and wind velocities defined over a 0.25$^\circ$x0.25$^\circ$ grid obtained from the fifth generation ECMWF reanalysis (ERA5, \citep{hersbach2023era5}).

To improve the quality of the forcing wind data we compared the ERA5 wind velocities to the IFREMER satellite radiometer velocities (https://cerweb.ifremer.fr/datarmor/products/satellite/l4/multi-sensor/ifr-l4-ewsb-blendedwind-glo-025-6h-rep/data/). The comparison subsequently lead to a bias correction of the ERA5 wind field.
This was performed since wind velocities provided from reanalyisis products are often biased at high winds (extreme conditions) and therefore a comparison with satellite radiometers (or observations) can result in a better treatment of the meteorological forcing \citep{campos2022assessment,benetazzo2022correction}. A similar approach was adopted by \cite{zhai2023applicability}, who performed a bias-correction of the ERA5 wind field in the SCS considering wind measurements from 15 buoys. The study revealed that ERA5 winds can underestimate or overestimate the wind intensities.

For the matter of this work, in order to correct the bias we preprocessed the reanalyis as follows.
The 10-meter northward and eastward wind velocity components provided by the ERA5 dataset span from 1970 to 2022 and are defined hourly on a regular grid of 0.25$^\circ$. In contrast, the IFREMER dataset contains wind velocity components from 1992 to 2020 that are available every 6 hours and with the same spatial resolution. Although the locations of the two datasets did not align, a series of pre-processing steps were undertaken to ensure consistent spatiotemporal resolution between them. First, a downsampling process was applied to the ERA5 velocities in order to convert them to a 6-hourly interval. Additionally, a cubic spatial interpolation was performed in order to account for the mismatched locations leading to the same spatiotemporal resolutions for both the datasets.
These initial steps were required to compute the 6-hourly biases for the years 1992-2020. For each year, the wind velocity measurements at each grid node were considered individually. 
The velocities occurring at each grid node, for a specific year, were grouped into bins of 1 m/s and the 6-hourly biases were divided into 0.10 m/s bins. Subsequently, a probability density function was computed enabling the assignment of velocities falling within a specific bin to the most probable 6-hourly bias bin value for that particular year. This last operation yielded a 6-hourly map representing the most likely biases for both the wind velocity components. To further fine-tune the biases, an additional cubic spatial interpolation was performed so that hourly maps of biases were generated. These maps were then added to the original ERA5 wind field, resulting in a corrected wind field suitable for utilization in the numerical simulations.

\subsection{Model validations and statistical performance indexes}\label{sec:Model validation and statistical performance indexes}
The validation of the model required at first the tuning of the integration time steps for both the SCHISM and the WWMIII models as well as the choice of the unstructured mesh size of which more details will be explained in section \ref{sec:Mesh Independence analysis}.
The overall quality of the results and the performance of the model were quantified through the comparison with observations provided by physical stations ($H_s$, $T_p$, $D_m$, $h$) and satellite measurements ($H_s$). \\

\subsubsection{Statistical analysis}\label{sec:statistical analysis}
The validation was carried out by evaluating a series of statistical indicators that are commonly used in literature \citep{shi201939,mentaschi2015performance} for assessing the accuracy of the results such as the Correlation coefficient $CC$,
Centered root mean square difference $RMSD$ and the Standard deviation $\sigma$.
These three fundamental indicators can be grouped within the Taylor diagram \citep{taylor2001summarizing}, which provides an effective graphical representation of the results.
As also mentioned by \cite{mentaschi2015performance} it is advised not to rely on the more traditional error indicators ( such as the Normalized Root Mean Square Error $NRMSE$ and the Scatter Index $SI$); instead, a more accurate evaluation of the errors can be performed with the exploitation of the symmetrically normalized root mean square error $HH$ \citep{hanna1985development,mentaschi2013problems}. \cite{mentaschi2013problems} pointed out that, in some circumstances, simulations that exhibit negative biases tend to be characterized by smaller $RMSE$ values: this implies that, paradoxically, these simulations perform better that simulations with absent biases. For these reasons, the statistical indicators considered in the validation for scalar integrated variables ($H_s$ and $T_p$) are: 

\begin{equation}
    \text{Correlation Coefficient}  \qquad \displaystyle CC = \frac{\sum (S_i - \bar{S}) \cdot (O_i - \bar{O})}{\sigma_S \cdot \sigma_O \cdot N} 
    \label{eqn:cc}
\end{equation}
\begin{equation}
    \text{Normalized Bias}  \qquad \displaystyle NBI = \frac{\sum (S_i - O_i)}{\sum O_i} \times 100
     \label{eqn:nbi}
\end{equation}
\begin{equation}
    \text{Hanna and Heinold indicator} \qquad \displaystyle HH = \sqrt{\frac{\sum (S_i - O_i)^2}{\sum (S_i \cdot O_i)}}
    \label{eqn:hh}
\end{equation}

$HH$ is also referred as a symmetrically normalized root mean square error: it takes into account both the average and scatter components of the error and it is unbiased towards simulations that underestimate the average. \\
In these formulations $N$ is the total number of $i\textsuperscript{th}$ samples, $S_i$ and $O_i$ are the simulated and observed variables with their respective means $\bar{S}$,$\bar{O}$ and standard deviations $\sigma_S$,$\sigma_O$.

A different treatment is reserved for $D_m$, which is a circular quantity, and the indicators analyzed for this matter were normalized with an angle of 2$\pi$ radiants \citep{mentaschi2015performance}: \\\\
\begin{equation}
    \text{Normalized Bias} \qquad \displaystyle  NBI_{D_m} = \frac{\sum \text{mod}_{-\pi,\pi}(S_i - O_i)}{2\pi N}
    \label{eqn:cc}
\end{equation}
\begin{equation}
    \text{Hanna and Heinold indicator} \qquad \displaystyle  HH_{D_m} = \sqrt{\frac{\sum\limits \left[\text{mod}_{-\pi,\pi}(S_i - O_i)\right]^2}{N}} \cdot \frac{1}{2\pi}
    \label{eqn:cc}
\end{equation}

where ${mod}_{-\pi,\pi}(S_i - O_i)$ is equal to $(S_i - O_i)-2\pi$ if $(S_i - O_i)>\pi$. Viceversa, ${mod}_{-\pi,\pi}(S_i - O_i)$ is equal to $(S_i - O_i)+2\pi$ if $(S_i - O_i)<-\pi$.
In order to better represent the accuracy of the comparisons, also the model skill score (skill) \citep{murphy1988skill}, the root mean square error  ($RMSE$) and the $BIAS$ were computed. \\

\begin{equation}
    \text{Model skill score} \qquad \displaystyle SKILL = \frac{\sum (O_i - S_i)^2}{\sum (O_i - \text{$\bar{O}$})^2}
    \label{eqn:cc}
\end{equation}
\begin{equation}
    \text{Root Mean Square Error} \qquad \displaystyle RMSE = \sqrt{\frac{\sum (O_i - S_i)^2}{N}}
    \label{eqn:cc}
\end{equation}
\begin{equation}
\text{BIAS} \qquad \displaystyle BIAS =\frac{\sum (S_i - O_i)}{N}
\label{eqn:cc}
\end{equation}

\subsubsection{Tidal and wave stations}\label{sec:Tidal and wave stations}
\begin{figure}[h!]
\begin{center}
\includegraphics[width=1\textwidth]{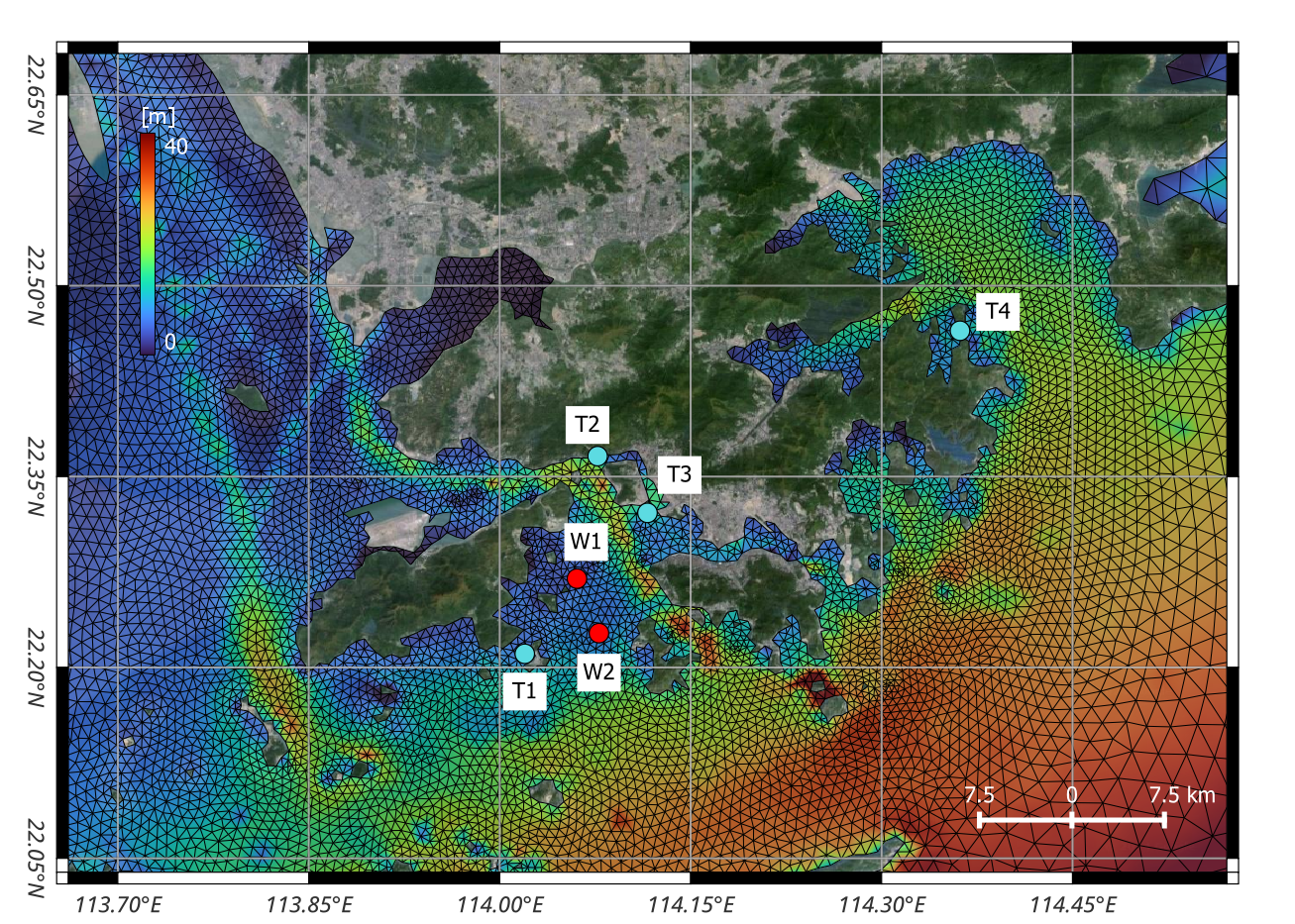}
\end{center}
\caption{Wave stations and tide stations considered for the model validation}\label{fig:wave stations and tide stations}
\end{figure}
The simulated water levels ($h$) and the simulated wave variables considered in this paper ($H_s$, $T_p$ and $D_m$) were compared to the observations provided in the stations shown in Figure \ref{fig:wave stations and tide stations}.
More specifically, the four tide stations T1, T2, T3 and T4, of which more details can be seen in table \ref{table:all_stations_infos}, are managed by the Hydrographic Office of Marine Department (HO) and provide water level values every 10 minutes. These records are referred to the Hong Kong time, which is 8 hours head of the Coordinated Universal Time UTC and their values are in meters above the Chart Datum , which is 0.146 meters above the Hong Kong Principal Datum. The two only existing wave stations in the territory of Hong Kong, W1 and W2 (details in table \ref{table:all_stations_infos}), are handled by the Port Works Division of the Hong Kong Civil Engineering and Development Department (CEDD). Specifically, they are bed-mounted wave recorders that provide a hourly long-term wave monitoring programme in the harbour starting from 1994.
\begin{table}[h!]
\resizebox{\textwidth}{!}{%
\begin{tabular}{cc|cccc}
\textbf{Station ID} & \textbf{Name}      & \textbf{WGS84 Longitude} & \textbf{WGS84 Latitude} & \textbf{Variables} & \textbf{Data availability} \\ \hline
W1                  & Kau Yi Chau        & 114.0644                 & 22.2649                 & wave data          & Jan 1994 - Dec 2022     \\
W2                  & West Lamma Channel & 114.0767                 & 22.2208                 & wave data          & Jan 1994 - Dec 2022     \\
T1                  & Cheung Chau        & 114.0231                 & 22.2142                 & water level        & Feb 2005 - Aug 2020     \\
T2                  & Ma Wan             & 114.0713                 & 22.3640                 & water level        & Dec 2004 - Aug 2020     \\
T3                  & Kwai Chung         & 114.1227                 & 22.3237                 & water level        & Sept 2001 - Aug 2020    \\
T4                  & Ko Lau Wan         & 114.3608                 & 22.4587                 & water level        & Jan 2000 - Aug 2020    
\end{tabular}%
}
\caption{Details of the stations used for the validation of the model}\label{table:all_stations_infos}
\end{table}

\subsubsection{Satellite wave heights}\label{sec:Satellite wave heights}
\begin{figure}[h!]
\begin{center}
\includegraphics[width=1\textwidth]{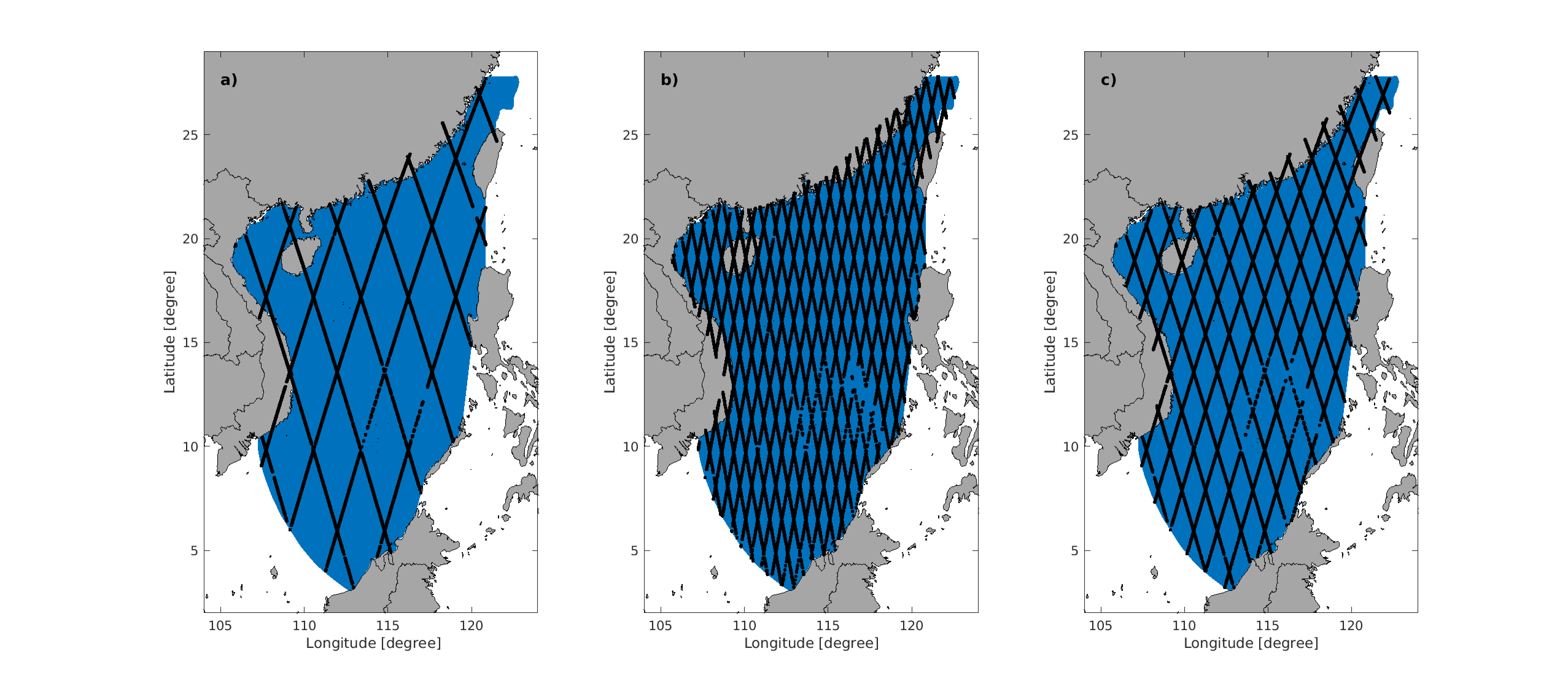}
\end{center}
\caption{Plots of satellite orbits for the three satellites considered: a) TOPEX 2003, b) ENVISAT 2004, c) JASON-2 2016}\label{fig:satellite tracks}
\end{figure}

Further validation of the simulated wave parameters was done using satellite altimetry.
The data is retrieved from the European Space Agency (ESA) Sea State Climate Change Initiative (CCI) project \citep{piolle2020esa,dodet2020sea} that produced a global dataset of $H_s$ with a spatial resolution of approximately 6 km for each of the satellites deployed in the missions . This paper considers the database version 1.1 ( Topex satellite from 1992 to 2005) and the database version 3 (Envisat satellite from 2002 to 2012 and Jason-2 satellite from 2008 to 2019): figure \ref{fig:satellite tracks} shows the satellite tracks in the area of interest.
The modeled $H_s$ were interpolated in space and time in order to match the resolution of the measurements along the satellite tracks.

\section{Results}\label{sec:Results and discussion}

\subsection{Mesh Independence analysis}\label{sec:Mesh Independence analysis}
One of the strictest requirements that any explicit numerical scheme has to fulfill in order to be stable and convergent is the so-called Courant-Friedrichs-Lewy ($CFL$) condition \cite{courant1928partiellen}: $CFL$ $<$ 1, where $CFL = \displaystyle\frac{\Delta t \cdot v}{\Delta x}$. In this formula, $\Delta t$ is the integration time step, $v$ is the flow velocity and $\Delta x$ is the mesh size. However, the fact that SCHISM is an unstructured model based on implicit time stepping schemes makes the $CFL$ constraint no longer a stringent restriction, as stressed by \cite{zhang2008selfe}. For this reason, even significant changes in both the mesh size and time step will not considerably affect the results. Several mesh sizes were tested, with minimum lengths ranging between 0.001$^\circ$ and 0.01$^\circ$ in the Hong Kong waters. In particular, five different meshes that differ for the resolution around Hong Kong but share the same grid size offshore and far from the area of interest were considered in the sensitivity analysis: mesh A (minimum size $~$ 0.003$^\circ$ ), mesh B (minimum size $~$ 0.004$^\circ$), mesh C (minimum size $~$ 0.006), mesh D (minimum size $~$ 0.008$^\circ$), mesh E (minimum size $~$ 0.01$^\circ$). 
\begin{figure}[h!]
\begin{center}
\includegraphics[width=1\textwidth]{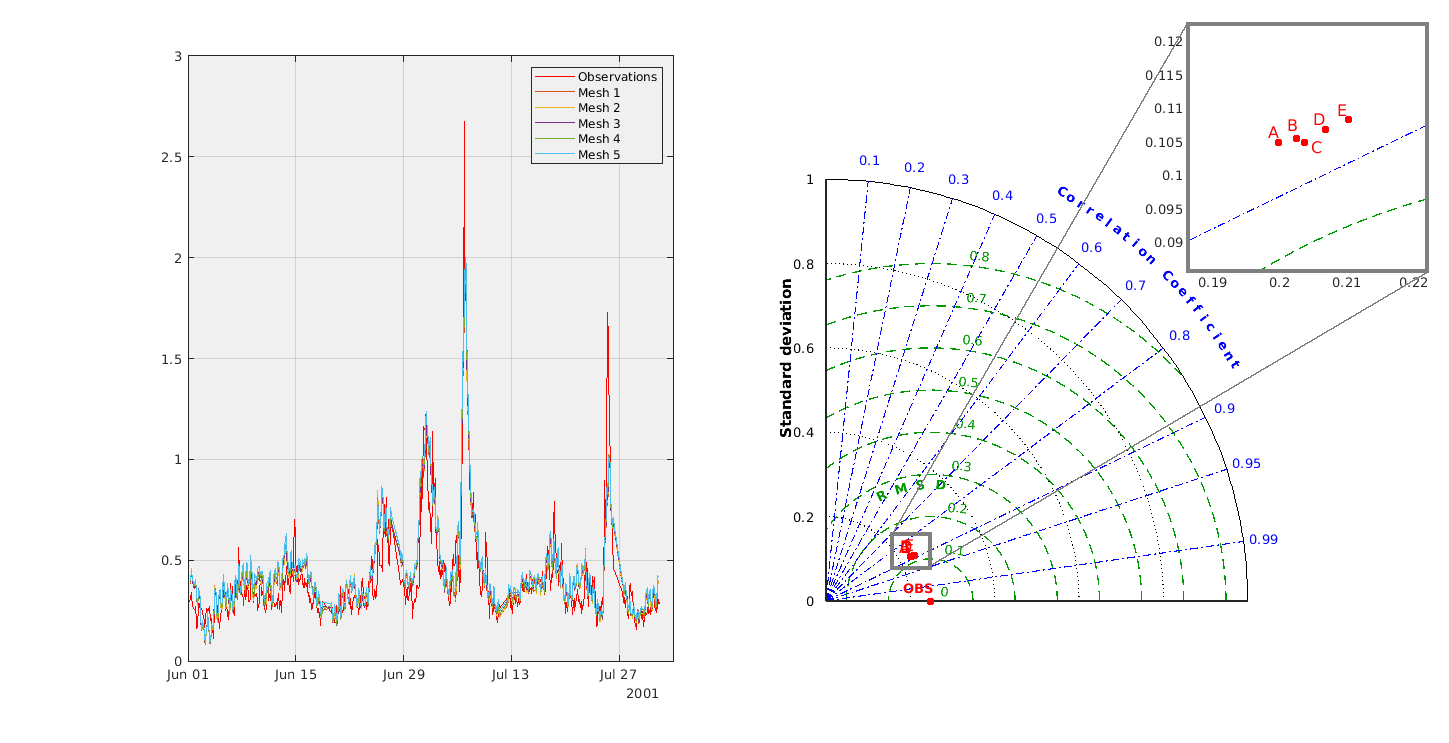}
\end{center}
\caption{Mesh sensitivity test on $H_s$ and related Taylor Diagram}\label{fig: sensitivity hs 2001}
\end{figure}

Figure \ref{fig: sensitivity hs 2001} illustrates the outcomes obtained for $H_s$ using the above mentioned unstructured meshes, along with the Taylor Diagram in W1 during a brief time period in 2001.
The correlation coefficient ($CC$) values remain consistently around 0.9, the root mean square difference ($RMSD$) hovers around 0.12 m, and the standard deviation ($STD$) remains approximately 0.22 m. These results show that increasing the nearshore mesh resolution more than 0.008$^\circ$ doesn't further improve the model skill.
For these reasons, a resolution of 0.008$^\circ$ was selected in order to provide sufficient resolution nearshore and reduce the computational time of the numerical simulations.

\subsection{ERA5 wind bias correction}\label{sec:ERA5 wind bias correction}
As anticipated in section \ref{sec:Study area and meteocean forcing data}, ERA5 wind velocities were compared to IFREMER wind velocities in order to apply a hourly bias correction to each node velocity and for the years spanning between 1992 and 2020. The mere comparison between the two datasets, both for the northward and eastward wind components, provides an understanding of the degree of similarity between their distributions. For this matter, due to the large domain area and the considerable amount of wind nodes, three different locations were selected and examined: location A (22$^\circ$N, 114$^\circ$E), location B (15$^\circ$N, 114$^\circ$E) and location C (11$^\circ$N, 109.25$^\circ$E). 
\begin{figure}[h!]
\begin{center}
\includegraphics[width=1\textwidth]{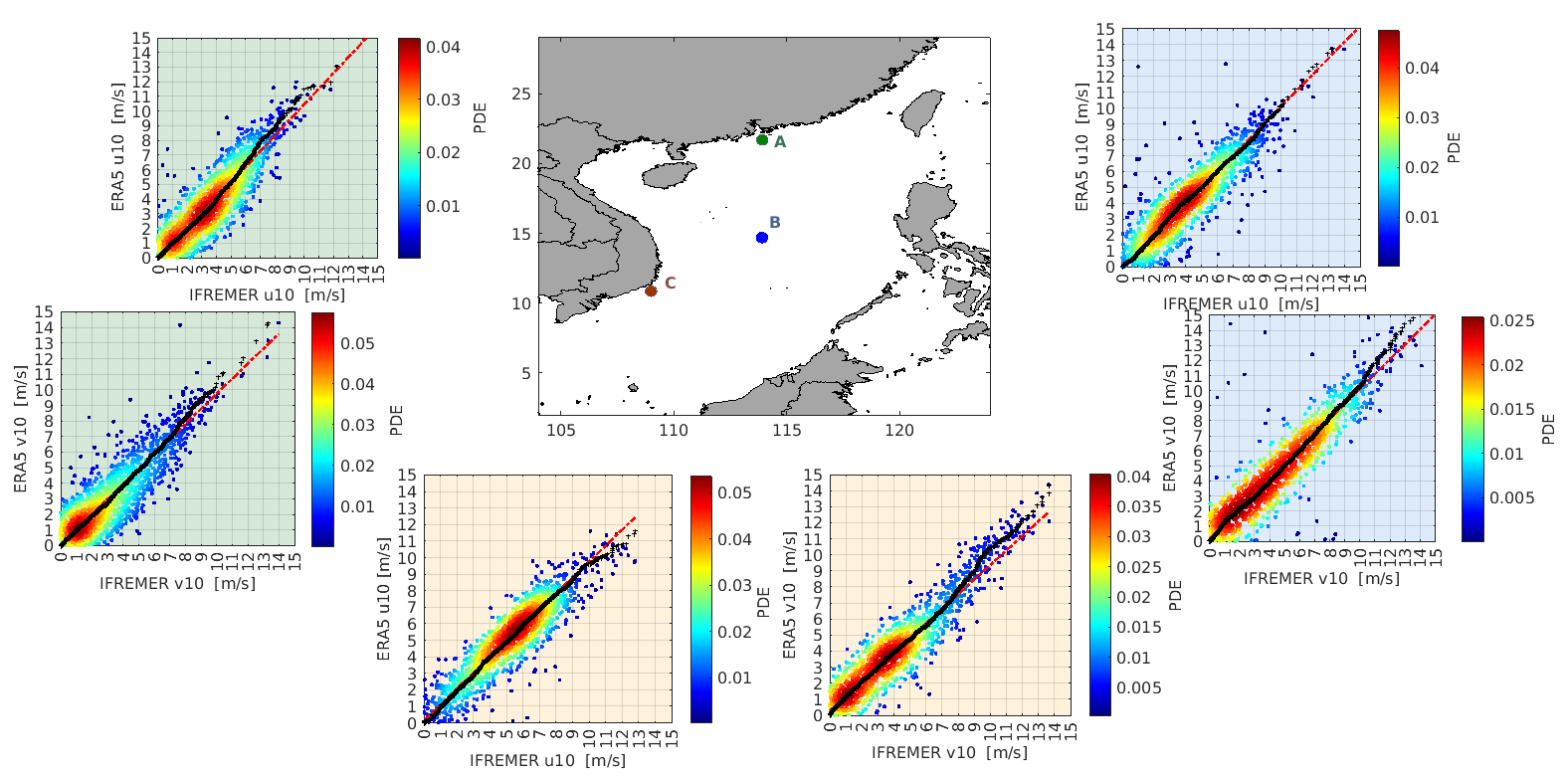}
\end{center}
\caption{Scatter plots and Q-Q plots related to the absolute values of wind velocities in locations A, B, C}
\label{fig:scatter plots wind}
\end{figure}
Figure \ref{fig:scatter plots wind} shows the positions of these three locations and the related  scatter plots where each mark is colored by the spatial density (PDE) of nearby points, computed by means of the Kernel smoothing function. The plots also display the quantiles of the IFREMER wind velocities versus the quantiles of the ERA5 wind velocities: in general, the two distributions are alike if they do not deviate from red dashed line. It can be observed that the two datasets show a strong agreement when the absolute value of their velocity does not exceed circa 7 m/s. In this range of velocities, the eastward and northward wind components therefore manifest robust values of $BIAS$,$RMSE$ and $CC$. The eastward wind component at Location A is characterized by $CC$ equal to 0.998, $RMSE$ equal to 0.169 m and a $BIAS$ of -0.025 m.
Beyond that threshold, the disparities between ERA5 and IFREMER, although still within the range of approximately  of 1 m/s, begin to grow. In this case, location A shows poorer statistic metrics ($CC$= 0.979, $RMSE$=1.207 m and $BIAS$=-1.073 m).
From the plots it seems that ERA5 velocities tend to be higher then IFREMER velocities for values greater than 7 m/s. However, location C seems to show an opposite behaviour when the absolute value of the eastward wind velocity exceeds the value of 10 m/s: $BIAS$ reaches a positive value of 0.117 m, $RMSE$ is equal to 0.126 m and $CC$ is equal to 0.983. Given the diverse nature of the domain , it is challenging to identify a distinct behavior that applies uniformly along the wind nodes. Yearly colored maps were generated in order to offer a glimpse of how the mean wind bias is distributed along the SCS: it is therefore possible to understand when the bias becomes positive and when it becomes negative. To enhance the visualization of the results and facilitate plotting, the colorbar range was adjusted between -1 m/s and 1 m/s. This range was selected based on the observation that the bias typically fluctuates within these values.
\begin{figure}[h!]
\begin{center}
\includegraphics[width=1\textwidth]{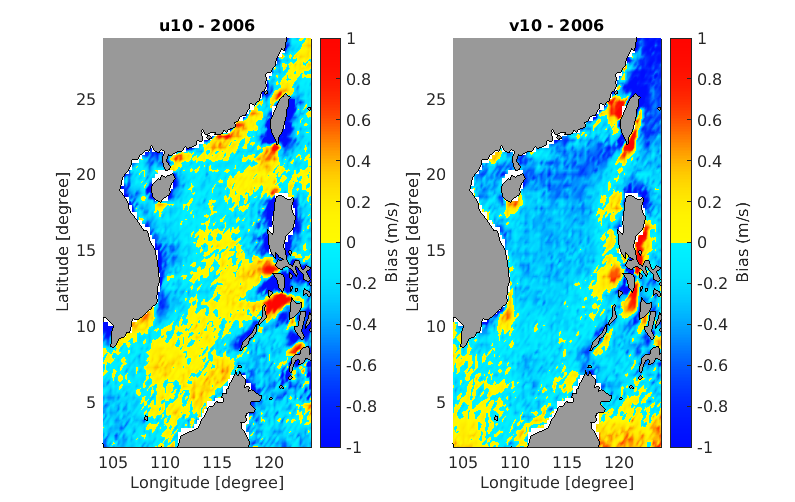}
\end{center}
\caption{Mean annual bias in 2006 for both eastward and northward wind velocity components}\label{fig:bias 2006}
\end{figure}

\begin{figure}[h!]
\begin{center}
\includegraphics[width=1\textwidth]{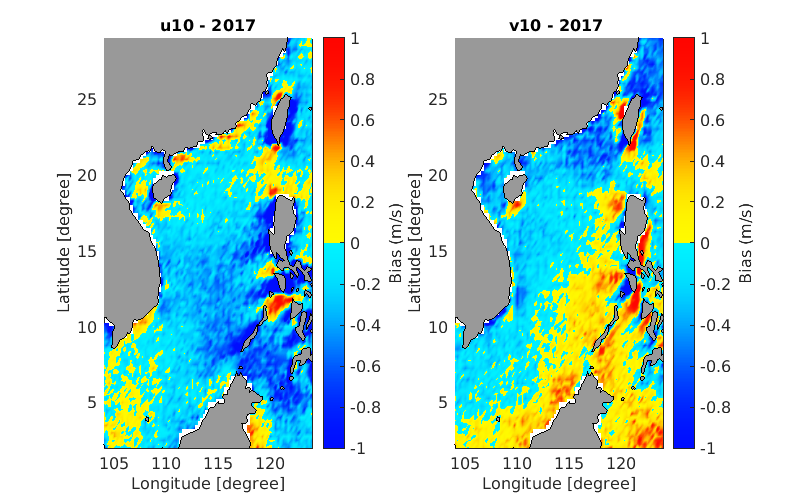}
\end{center}
\caption{Mean annual bias in 2017 for both eastward and northward wind velocity components}\label{fig:bias 2017}
\end{figure}

Figures \ref{fig:bias 2006} and \ref{fig:bias 2017} show the mean of the bias for the two velocity components for the years 2006 and 2017. In 2006, the observational data for eastward wind velocities ($u_{10}$) in the coastal regions between Hainan Island and Taiwan, Taiwan and the northern part of the Philippines, Manila and north of Brunei, and Brunei and Ho Chi Minh City showed positive biases. Differently, the northward wind velocities ($v_{10}$) are predominantly negative with a few positive patches around Taiwan, at the Philippines and in proximity of Hainan Island. In the year 2017, there appears to be a noticeable resemblance in the behavior of both $u_{10}$ and $v_{10}$ mean biases. Similar to the biases observed in 2006, the coastline extending from the northern open boundary to the left edge of the southern open boundary within the numerical domain exhibits comparable biases. However, in the eastern region of the domain, certain areas around the Philippines, and generally in deep water locations, the mean bias for both components exhibits an opposite sign.

\subsection{Model performance}\label{sec:Wave results and model performance}
The overall performance of the SCHISM-WWMIII two-way coupled model deployed in this study was assessed by considering and analyzing the water level elevations $h$, significant wave heights $H_s$, peak wave period $T_p$ and mean wave direction $D_m$. 

As mentioned in section \ref{sec:Tidal and wave stations}, 4 tide stations in Hong Kong (T1, T2, T3, T4) handled by the Hydrographic Office of Marine Department (HO) were considered for the water level validation. We compared the hourly modeled water elevation versus hourly observed water elevation for each of the tide stations and for the entire simulation time (1970-2022). Generally, all the locations exhibit similar behavior and demonstrate a solid agreement with the observations, with the exception of a few observed negative water level values of which the respective simulated values appear to be positive. 
\begin{table}[]
\begin{tabular}{c|cccl}
\textbf{Station ID} & \textbf{CC} & \textbf{BIAS [m]} & \textbf{RMSE [m]} & \textbf{skill} \\ \hline
T1                  & 0.845       & 0.081         & 0.369         & 0.742          \\
T2                  & 0.839       & 0.102         & 0.379         & 0.818          \\
T3                  & 0.803       & 0.085         & 0.407         & 0.720          \\
T4                  & 0.780       & 0.144         & 0.407         & 0.717         
\end{tabular}%
\caption{Statistics related to the water levels}\label{table:statistics_tide_levels}
\end{table}
Table \ref{table:statistics_tide_levels} summarizes the values of the statistical parameters evaluated to assess the performance of the comparisons. All the four stations are depicted by  $CC$ values in the range 0.780 (for T4) and 0.845 (for T1), $BIAS$ values of the order of 0.1 m and $RMSE$ values varying between 0.369 m and 0.407 m. The $SKILL$ parameters for all the locations are greater than 0.7, with the highest value of 0.818 for T2, which suggest a robust agreement between the observations and the simulated variables. 
\begin{figure}[h!]
\begin{center}
\includegraphics[width=1\textwidth]{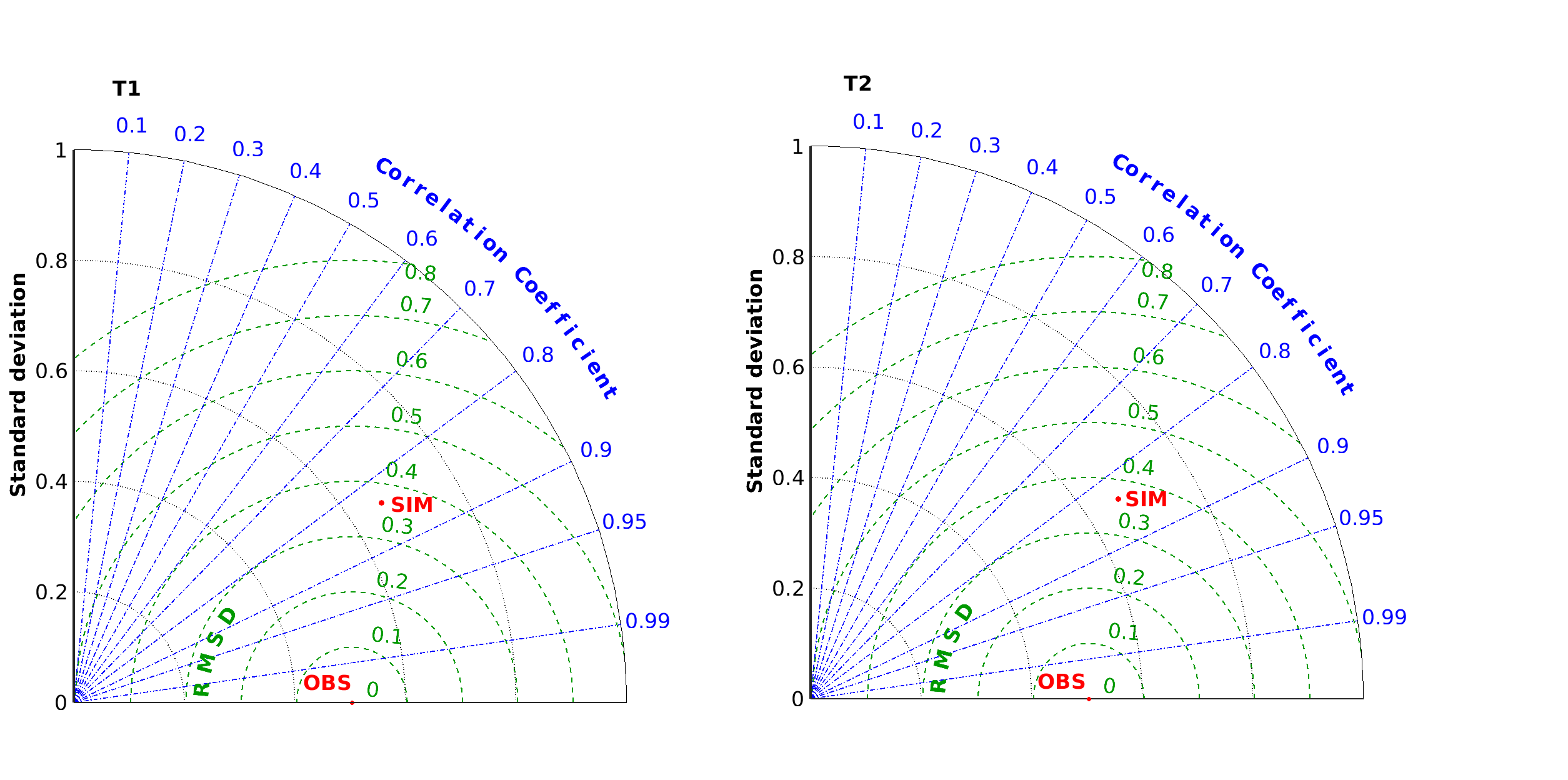}
\includegraphics[width=1\textwidth]{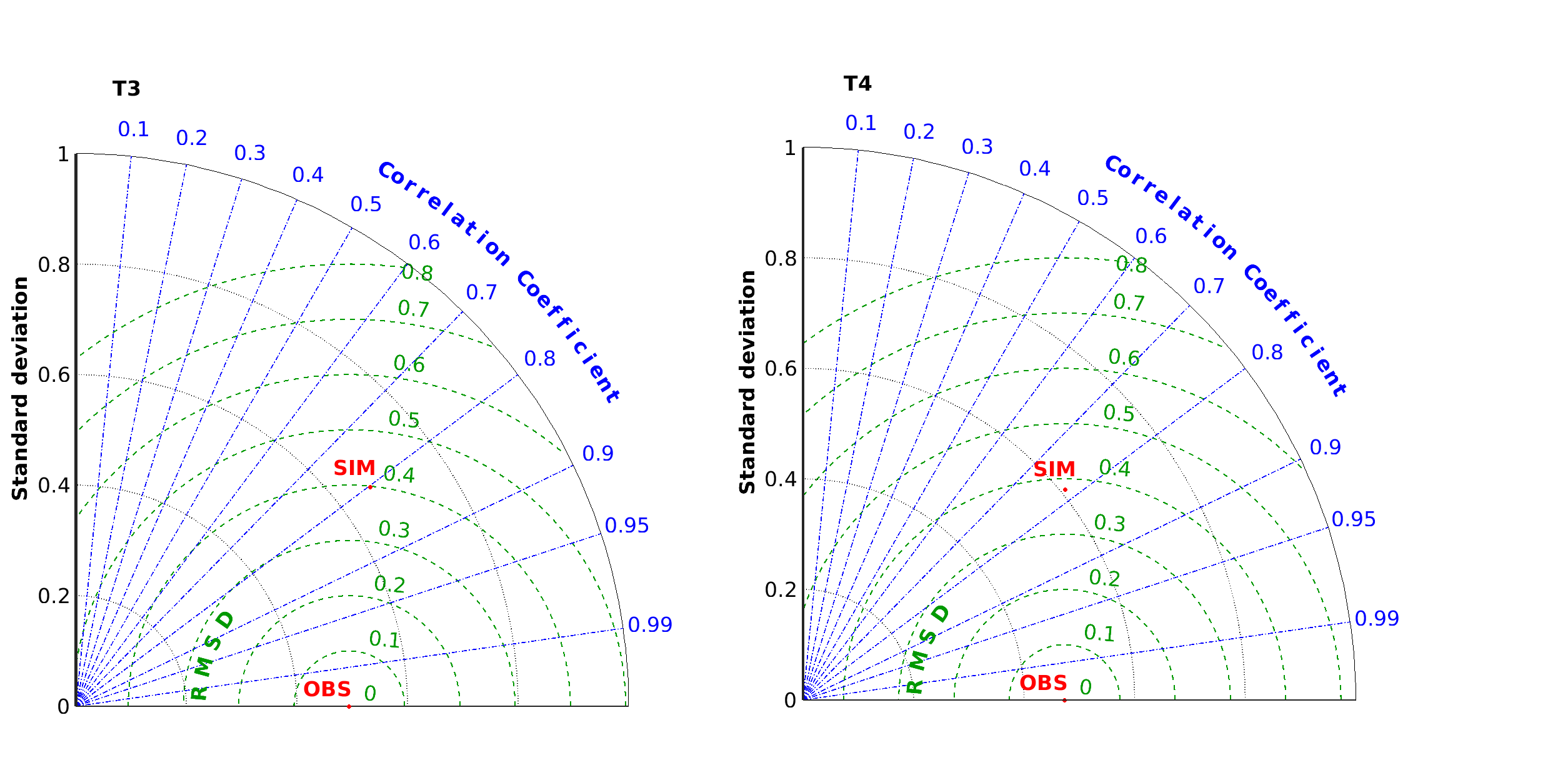}
\end{center}
\caption{Taylor diagrams related to the four tide stations}\label{fig:Taylor all tide stations}
\end{figure}
Figure \ref{fig:Taylor all tide stations} displays the Taylor diagrams for all the tide stations in order to provide more information about the comparison. Here, it is possible to see that the $RMSD$ values are always less than 0.4 m and that the standard deviations of the simulated water levels $h$ fluctuate around 0.6 m.

Regarding the wave heights validation, we started comparing the numerical model results with the satellite observations. in particular, the significant wave height $H_S$ for the whole numerical domain were interpolated along the satellite tracks, as described in \ref{sec:Satellite wave heights}.
\begin{table}[]
\begin{tabular}{c|ccccc}
\textbf{Year} & \textbf{CC} & \textbf{BIAS [m]} & \textbf{RMSE [m]} & \textbf{NBI [\%]} & \textbf{HH} \\ \hline
\textbf{1992} & 0.884       & -0.011        & 0.475                          & -0.50      & 0.204       \\
\textbf{1993} & 0.925       & -0.060        & 0.403                          & -3.90      & 0.225       \\
\textbf{1994} & 0.908       & -0.096        & 0.391                          & -6.44      & 0.234       \\
\textbf{1995} & 0.935       & -0.068        & 0.384                          & -4.34      & 0.208       \\
\textbf{1996} & 0.906       & -0.091        & 0.441                          & -5.70      & 0.244       \\
\textbf{1997} & 0.923       & -0.054        & 0.347                          & -3.89      & 0.217       \\
\textbf{1998} & 0.935       & -0.059        & 0.350                          & -4.44      & 0.219       \\
\textbf{1999} & 0.930       & -0.019        & 0.417                          & -1.17      & 0.214       \\
\textbf{2000} & 0.930       & -0.038        & 0.371                          & -2.41      & 0.204       \\
\textbf{2001} & 0.918       & -0.063        & 0.381                          & -4.11      & 0.219       \\
\textbf{2002} & 0.921       & -0.055        & 0.341                          & -3.92      & 0.213       \\
\textbf{2003} & 0.929       & -0.072        & 0.371                          & -4.65      & 0.210       \\
\textbf{2004} & 0.907       & -0.060        & 0.393                          & -3.86      & 0.225       \\
\textbf{2005} & 0.899       & -0.103        & 0.375                          & -7.82      & 0.253      
\end{tabular}%
\caption{TOPEX Statistics 1992-2005}\label{table:TOPEX}
\end{table}

\begin{table}[]
\begin{tabular}{c|ccccc}
\textbf{Year} & \textbf{CC} & \textbf{BIAS [m]} & \textbf{RMSE [m]} & \textbf{NBI [\%]} & \textbf{HH} \\ \hline
\textbf{2002} & 0.923       & -0.074        & 0.363                          & -4.86      & 0.212       \\
\textbf{2003} & 0.921       & -0.074        & 0.372                          & -5.01      & 0.220       \\
\textbf{2004} & 0.914       & -0.043        & 0.362                          & -2.92      & 0.218       \\
\textbf{2005} & 0.922       & -0.091        & 0.391                          & -6.00      & 0.225       \\
\textbf{2006} & 0.922       & -0.066        & 0.394                          & -4.32      & 0.224       \\
\textbf{2007} & 0.935       & -0.055        & 0.374                          & -3.59      & 0.208       \\
\textbf{2008} & 0.928       & -0.035        & 0.384                          & -2.25      & 0.211       \\
\textbf{2009} & 0.931       & -0.024        & 0.359                          & -1.54      & 0.201       \\
\textbf{2010} & 0.920       & -0.126        & 0.387                          & -9.04      & 0.244       \\
\textbf{2011} & 0.940       & -0.035        & 0.373                          & -2.06      & 0.190       \\
\textbf{2012} & 0.921       & -0.008        & 0.385                          & -0.49      & 0.198      
\end{tabular}%
\caption{ENVISAT Statistics 2002-2012}\label{table:ENVISAT}
\end{table}

\begin{table}[]
\begin{tabular}{c|ccccc}
\textbf{Year} & \textbf{CC} & \textbf{BIAS [m]} & \textbf{RMSE [m]} & \textbf{NBI [\%]} & \textbf{HH} \\ \hline
\textbf{2008} & 0.923       & -0.037        & 0.399                          & -2.37      & 0.218       \\
\textbf{2009} & 0.921       & -0.025        & 0.379                          & -1.58      & 0.208       \\
\textbf{2010} & 0.924       & -0.101        & 0.371                          & -7.32      & 0.233       \\
\textbf{2011} & 0.941       & -0.022        & 0.372                          & -1.31      & 0.189       \\
\textbf{2012} & 0.913       & -0.067        & 0.370                          & -4.39      & 0.217       \\
\textbf{2013} & 0.922       & -0.065        & 0.375                          & -4.10      & 0.209       \\
\textbf{2014} & 0.930       & -0.034        & 0.367                          & -2.31      & 0.214       \\
\textbf{2015} & 0.935       & -0.054        & 0.338                          & -3.86      & 0.208       \\
\textbf{2016} & 0.940       & -0.044        & 0.351                          & -3.01      & 0.203       \\
\textbf{2017} & 0.934       & -0.047        & 0.392                          & -2.74      & 0.199       \\
\textbf{2018} & 0.931       & 0.012         & 0.346                          & 0.82       & 0.199       \\
\textbf{2019} & 0.912       & -0.009        & 0.315                          & -0.64      & 0.201      
\end{tabular}%
\caption{JASON-2 Statistics 2008-2019}\label{table:JASON-2}
\end{table}

Tables \ref{table:TOPEX},\ref{table:ENVISAT} and \ref{table:JASON-2} show the findings of the yearly statistical analysis for each of the satellite missions considered. It can be observed that the correlation between satellite data and the simulation is significantly high with $CC$ values between 0.88 (TOPEX satellite - year 1992) and 0.94 (JASON-2 satellite - year 2011). The biases always appear to be negative with a minimum of -0.008 (ENVISAT satellite - 2012) except for a positive bias of 0.012 (JASON-2 satellite - year 2018). The $RMSE$  values show relatively low values and don't exceed 0.475 m. The normalized bias ($NBI$) parameters, which serve as indicators of the average component of the error, manifest minimal values. And the same can be said for the Hanna and Heinold ($HH$) indicators that oscillate around 0.2 therefore suggesting good quality of the comparisons. 

\begin{table}[]
\begin{tabular}{cc|cccccc}
\textbf{Mission} & \textbf{Coverage} & \textbf{CC} & \textbf{BIAS [m]} & \textbf{RMSE [m]} & \textbf{NBI [\%]} & \textbf{HH} \\ \hline
TOPEX            & 1992-2005         & 0.922       & -0.063        & 0.385                          & -4.16      & 0.220       \\
ENVISAT          & 2002-2012         & 0.926       & -0.060        & 0.377                          & -3.92      & 0.214       \\
JASON-2          & 2008-2019         & 0.929       & -0.043        & 0.365                          & -2.82      & 0.208      
\end{tabular}%
\caption{Statistics related to all the satellite missions}\label{table:all_satellite}
\end{table}

\begin{figure}[h!]
\begin{center}
\includegraphics[width=1\textwidth]{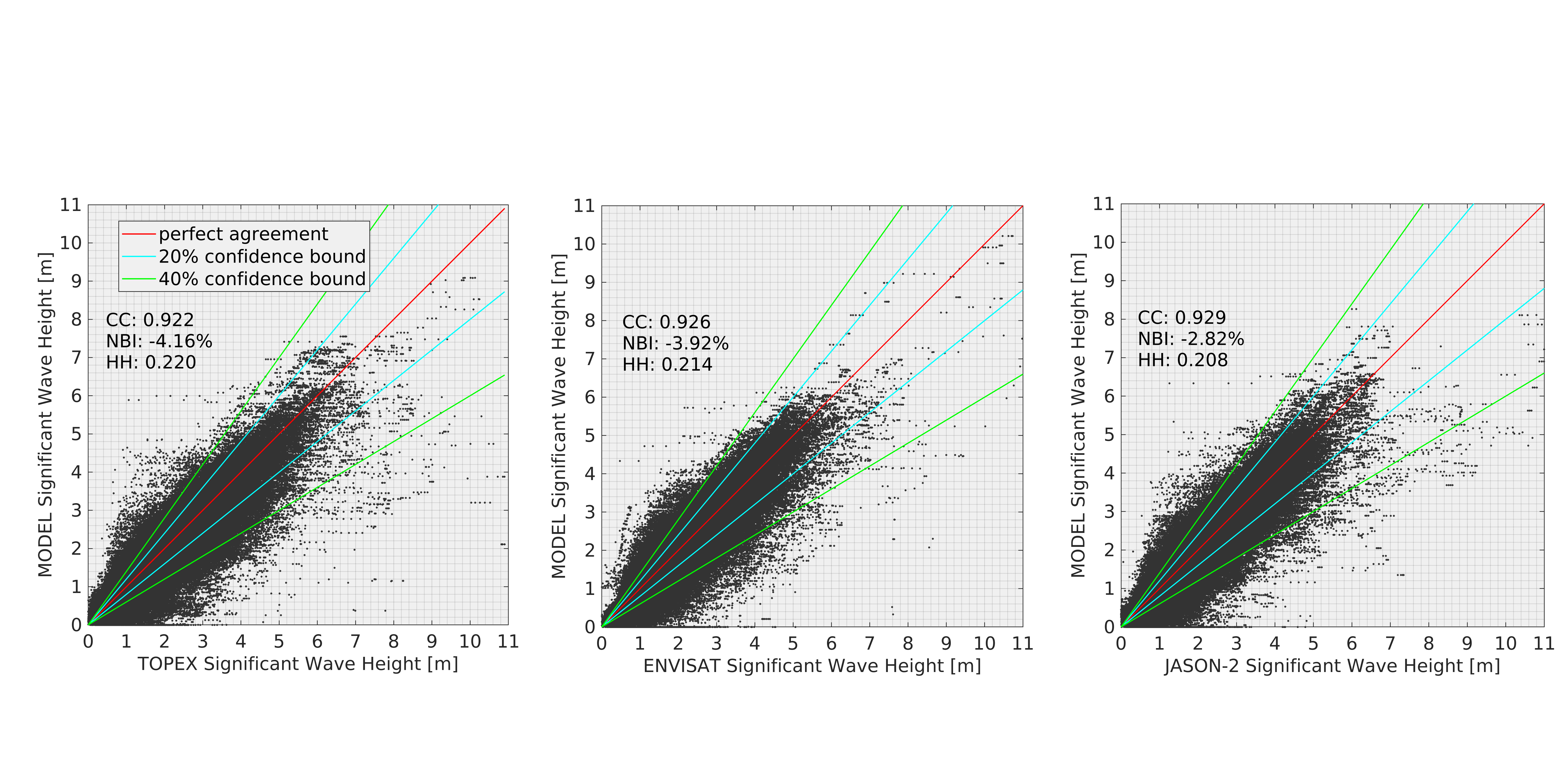}
\end{center}
\caption{Scatter plots of $H_s$ related to the three satellite missions}\label{fig:scatter plots Hs satellites}
\end{figure}

The statistical parameters were also evaluated for the whole satellite missions, as seen in table \ref{table:all_satellite} and from the scatter plots of Figure \ref{fig:scatter plots Hs satellites}. The overall good performance of the present hindcast when comparing the significant wave heights to the satellite observations is confirmed considering the whole observation periods. 
Notably, the bias and, consequently the normalized bias, are on average always negative, indicating that the numerical model tends to underestimate the wave height predictions.

Regarding the in-situ comparisons, the historical data provided by the two bed-mounted wave recorders in Hong Kong (W1 and W2) managed by the CEDD Department were utilized in order to validate the model and to check the quality of the results. The exact coordinates and other additional details can be seen in table \ref{table:all_stations_infos} and \ref{fig:wave stations and tide stations}.

\begin{figure}[h!]
\centerline{
\includegraphics[width=1\textwidth]{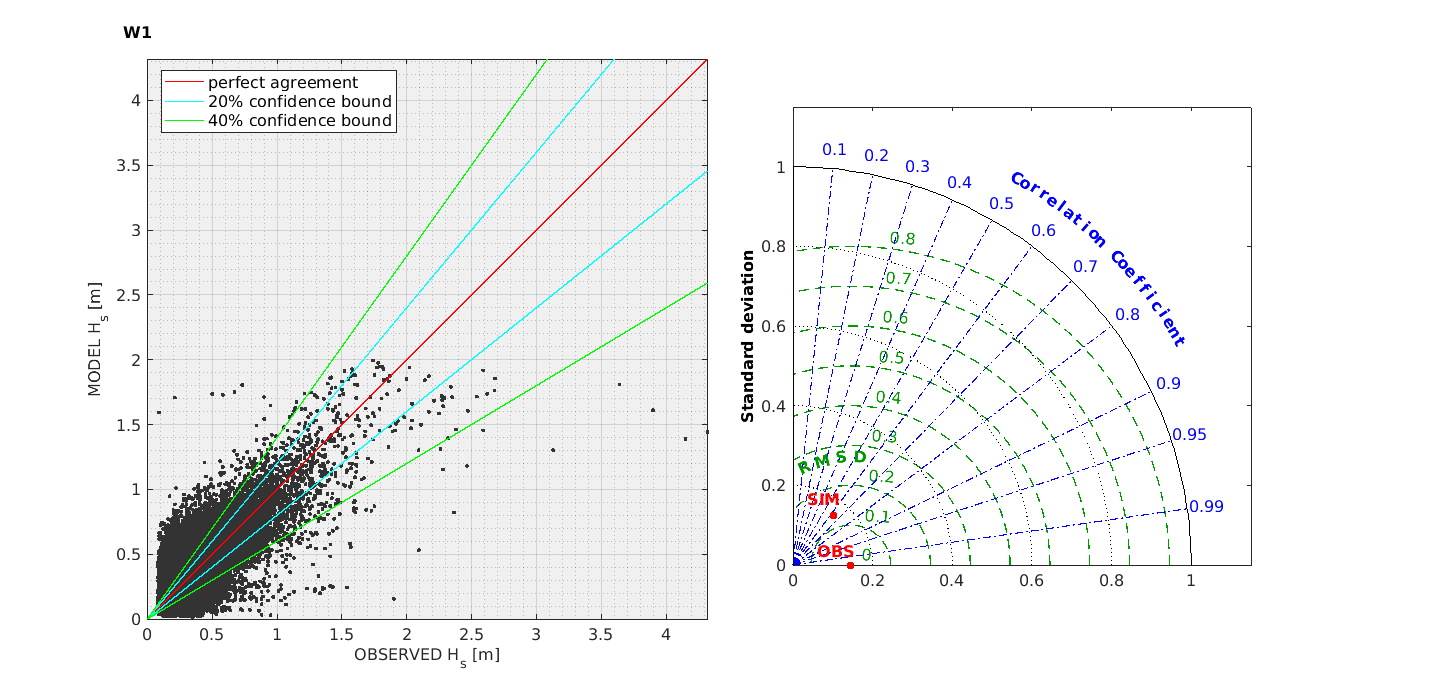}}
\centerline{
\includegraphics[width=1\textwidth]{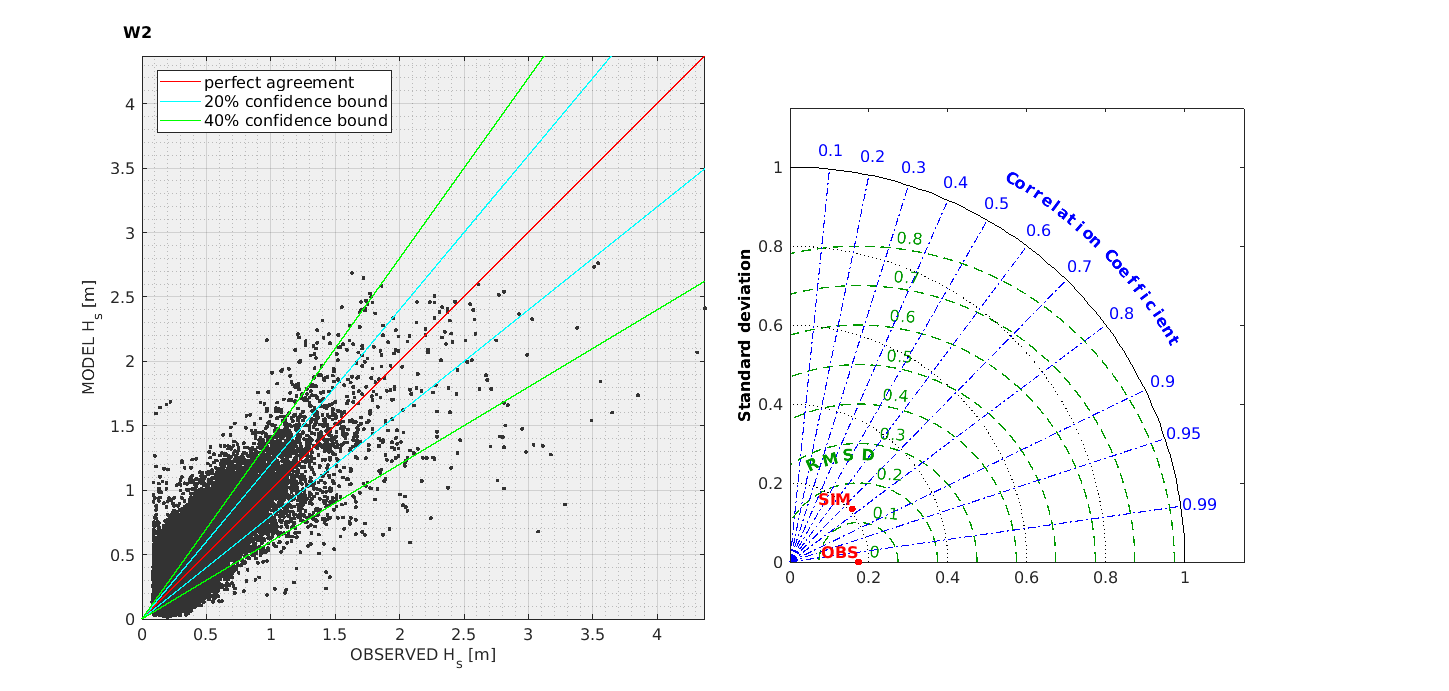}
}
\caption{Scatter plots and Taylor Diagrams related to $H_s$ for W1 (first row) and station w2 (second row)}\label{fig:HS scatter and Taylor}
\end{figure}

The quality assessment was performed by comparing the significant wave height ($H_s$), peak wave period ($T_p$) and mean wave direction ($D_m$) obtained by the wave station to the simulated ones and by computing a statistical analysis. Both W1 and W2 observed significant wave heights ($H_s$), as it can be seen from the scatter plots and Taylor diagrams in figure \ref{fig:HS scatter and Taylor} for both stations, seem to well represent the simulation findings. 

Results related to W1 and W2 generally show good agreements for $H_s$ values greater than 1 m with only a few outliers. Significant wave heights smaller than 1 m tend to be overestimated by the here deployed numerical model. The complexity of the coastline and the bathymetry around the two investigated sites could be the reasons why the comparison is not optimal. The Taylor diagrams show values of $RMSD$ around 0.15 and standard deviation around 0.2 for both stations while the correlation coefficients $CC$ reach values of  0.64 and 0.76 for W1 and W2 respectively. More details can be seen in table \ref{table:statistics_HS_W1_W2}.
\begin{table}[]
\begin{tabular}{cc|ccccc}
\textbf{Station ID} & \textbf{Variable} & \textbf{CC} & \textbf{BIAS} & \textbf{RMSE} & \textbf{NBI [\%]} & \textbf{HH} \\ \hline
W1                  & $H_s$             & 0.637       & 0.038 m        & 0.136 m         & 12.32       & 0.391       \\
                    & $T_p$             & 0.444       & 0.049 s        & 2.218 s        & 0.83       & 0.365       \\
W2                  & $H_s$             & 0.761       & 0.067 m        & 0.151 m        & 18.40       & 0.353       \\
                    & $T_p$             & 0.499       & -0.038 s       & 2.169 s        & -0.57      & 0.318      
\end{tabular}%
\caption{Statistics related to significant wave heights and peak wave periods for W1 and W2}\label{table:statistics_HS_W1_W2}
\end{table}
In the same table, the statistical parameters referred to the peak wave periods ($T_p$) are reported. Correlation values $CC$ for both W1 and W2 are considerably lower than the values obtained for the significant wave heights ($H_s$). $RMSD$ values are more or less equal to 2s and the standard deviations of the simulated peak wave periods ($T_p$) are a little less than 2s.

\begin{figure}[h!]
\begin{center}
\includegraphics[width=1\textwidth]{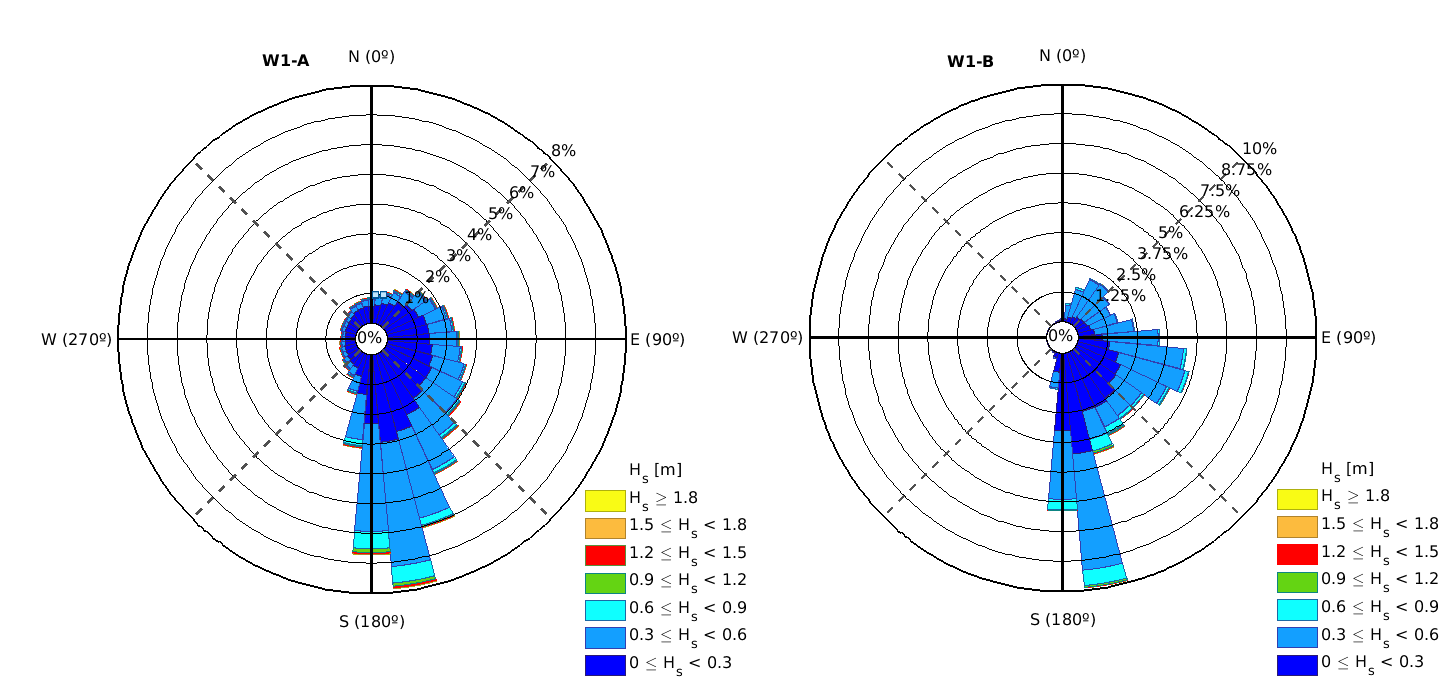}
\includegraphics[width=1\textwidth]{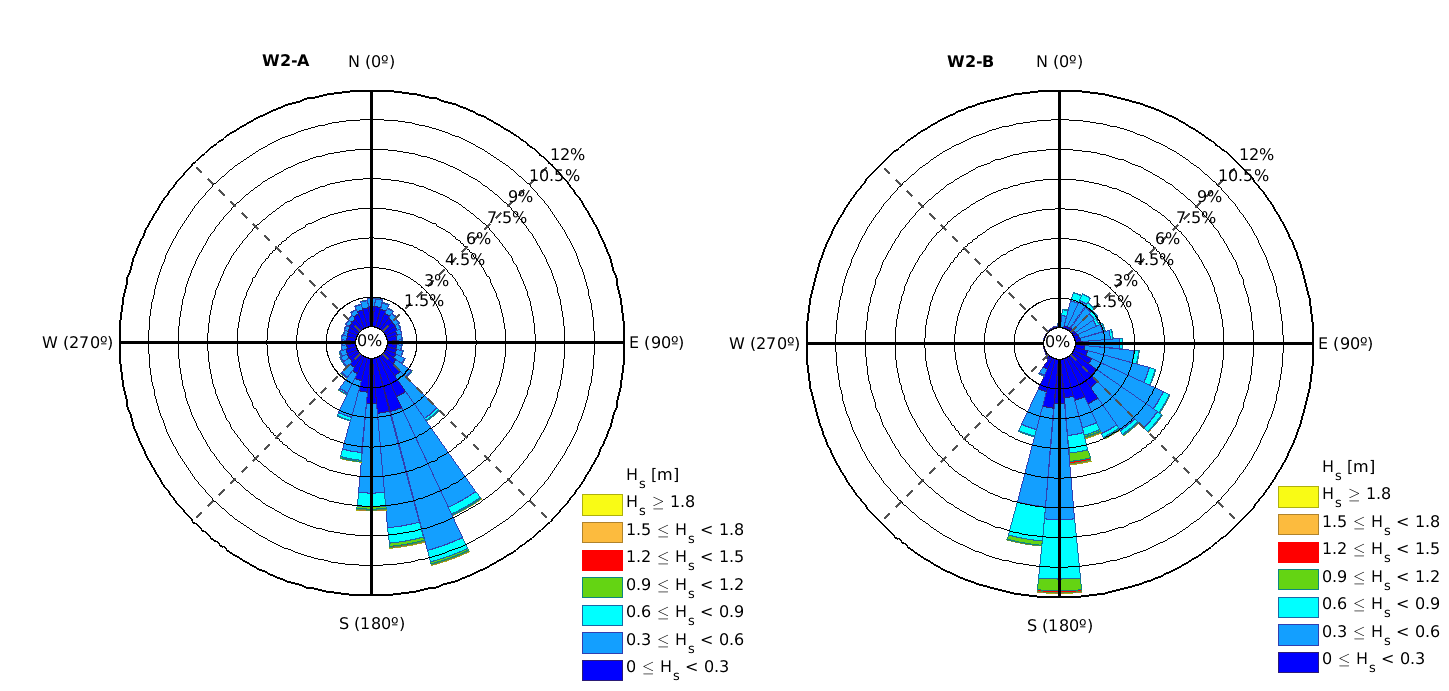}
\end{center}
\caption{Wave roses for both W1 and W2. The plots denoted by A) show the observed $D_m$ and the related $H_s$ while the B) plots are pertinent to the simulations}\label{fig:wind roses}
\end{figure}
The comparative analysis of mean wave directions ($D_m$) was conducted using wave roses, as depicted in Figure \ref{fig:wind roses}. These visual representations illustrate the average directions from which the waves originate, along with their corresponding significant wave heights ($H_s$). As a result of the geographical positioning of the two wave stations, waves originating from the second quadrant (between North and West) are not captured in the simulations. Nevertheless, the observed series of mean wave directions for both stations exhibit the occurrence of minor waves emanating from that direction. This occurrence is likely attributed to the influence of marine traffic within the Hong Kong harbor, including boat wakes, as well as potential errors originating from the wave stations. The analysis reveals that in station W1, the most frequently occurring mean wave direction aligns well with the simulated results from the numerical model. However, this congruence does not hold true for station W2. In W2, the most frequently observed mean wave direction is South-East, whereas the most frequently simulated mean wave direction tends to be South/South-West. The statistical analysis on the whole dataset for W1 showed a $HH_{D_m}$ value of 0.289 and a $NBI_{D_m}$ equal to 0.250. The W2 station show almost the exact values: $HH_{D_m}$ equal to 0.288 and $NBI_{D_m}$ equal to 0.250. The $D_m$ results show a relatively poor agreement between the outputs of the numerical model and the observations as also highlighted by \cite{kong2024model}. \cite{kong2024model} compared the outcomes of the Operational Marine Forecasting System (OMFS) managed by the Hong Kong Observatory to the observations in both W1 and W2, stressing the fact that apparent discrepancies arise.

\section{Discussion}

\subsection{Tidal levels}
As shown in the previous section, the simulated water levels extracted at the four tidal stations in Hong Kong (T1,T2,T3,T4) show a good agreement when compared to the in-situ data. In fact, $CC$ values range between 0.780 and 0.845, $BIAS$ values are of the order of 0.1 m and $RMSE$ values range between 0.369 m and 0.407 m. 
\begin{figure}[h!]
\begin{center}
\includegraphics[width=1\textwidth]{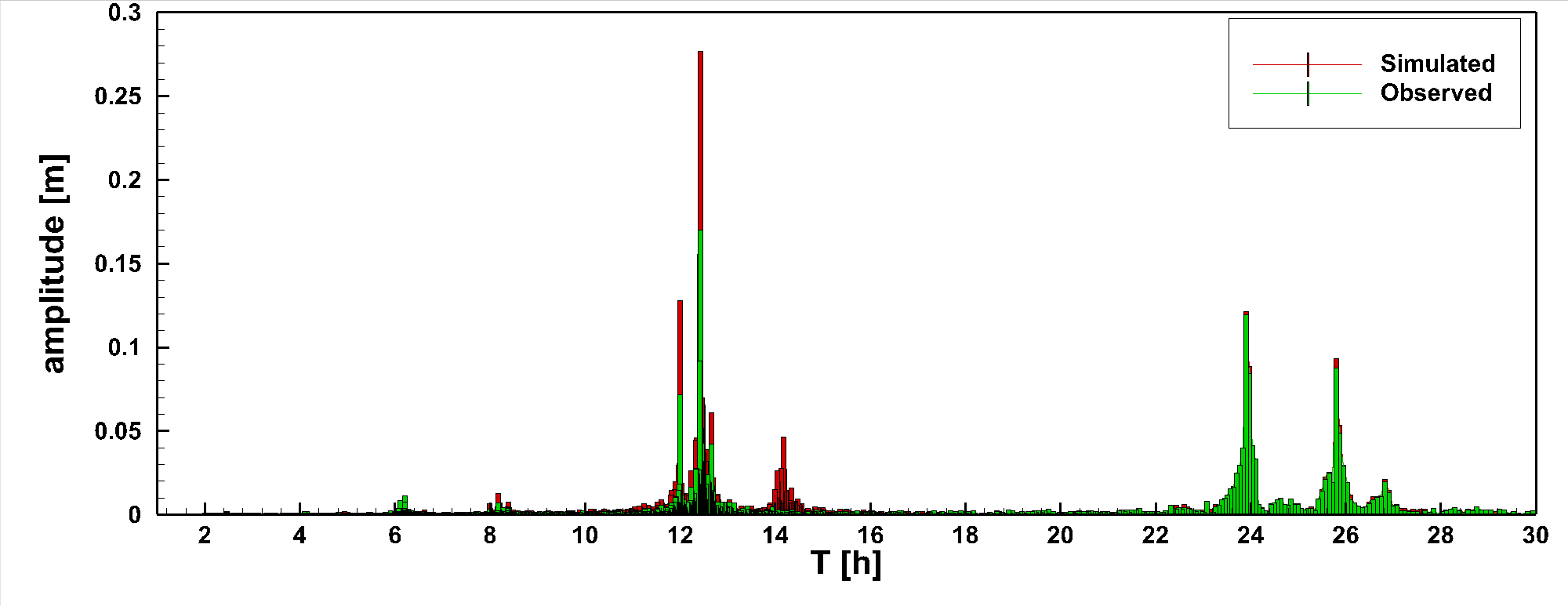}
\end{center}
\caption{Comparison of the Power Spectral Density functions of one-year tidal level signals computed with the numerical results and the in-situ observations, station ID T2.}\label{fig:psd}
\end{figure}
We also performed an harmonic analysis of the water levels and we compared it with the observations. Figure \ref{fig:psd} shows the Power Spectral Density function for one station, i.e. station T2, as representative of the general behaviour. it can be noted that the diurnal constituents are very well captured by the model, as the shallow water (less intense) constituents with periods around 6 and 8 hours. The larger difference can be observed for the diurnal constituents, especially for the M2 tides, which is the dominant. Most of the differences between the simulations and the observation can be attributed to this. The overestimation of the most intense tidal components could be attributed to the difficulty of the 2D depth averaged models in correctly represent the frictional dissipation compared to more accurate 3D models, even hydrostatic \citep{rozendaal2024relationship}.

In fact, comparing the performance of the present model with other global studies \citep{zhang2023global} or regional studies \cite{pan2020channel,he2022coastal}, we noticed that tidal levels are predicted with lower accuracy. Note that the mentioned study were based on three dimensional hydrodynamic models. The comparison and consequently the statistical metrics are strongly influenced by the resolution of the mesh, the details of the bathymetry and 3D baroclinic effects \citep{huang2022tidal}. Regarding regional studies, \cite{pan2020channel} and \cite{he2022coastal} computed the water level validations at the same locations deploying a 3D approach, increased resolution in shallow water areas and temperature and salinity as additional input fields. In this way they improved the performance of the model leading to $CC$ values ranging between 0.95 and 0.98, $RMSE$ values equal to approximately 0.10 m at all the tide stations and $SKILL$ values between 0.97 and 0.99. 

\subsection{Simulations of extreme events}
The performance of large scale wave hindcast are known to underperform in the case of extreme events, typically underestimating the significant wave heights \citep{mentaschi2023global}. This tendency is commonly ascribed to the reanalysis of the atmospheric forcing that tend to predict lower winds and atmospheric pressure extremes, typical of tropical storms or typhoons \citep{schenkel2012examination,hodges2017well,campos2022assessment,lodise2024performance}. We acknowledge the limitation of the present hindcast analysis to correctly reproduce wave fields generated by extreme events such as Tropical Storms and Typhoons, which would require a dedicated model set up. In particular, several approaches have been showed to effectively reproduce wave and storm surges caused by tropical extreme events \citep{holland1980analytic,emanuel2011self,yang2019comparative,wang2021numerical}. 
In the context of extreme events, operational wave models have been observed to underestimate the peak of extreme wave heights, with errors reaching several meters. This phenomenon is frequently observed in comparisons of model forecasts and hindcast to observations during the passage of cyclones \citep{cardone1996evaluation,collins2021altimeter}.A poor understanding of the physics that govern extreme regimes, the accuracy of wind forcing data, and the challenge of representing the aforementioned dynamics on discrete grids in time and space are often advocated as the main source of discrepancies. These factors lead to the smoothing and underestimation of maximum values, sharp gradients, and overall variability \citep{cavaleri2009wave}. This is particularly the case with regard to the area of highest waves, which can be observed in the vicinity of a cyclone's forward motion. \citep{collins2021altimeter}.

In order to discuss the performance of the present hindcast in simulating extreme events, we focused on three periods recorded by the wave buoys in the Hong Kong waters during the typhoon seasons for the years 2003, 2008 and 2018. In particular, we used the recordings from W2 (see Figure \ref{fig:wave stations and tide stations}) that is more exposed to the open ocean waves.  

In the following, we briefly describe the extreme events simulated, providing the main characteristics of the tropical storms and typhoons. The data have been collected by the official report yearly issued by the Hong Kong Observatory, which is a government department responsible for monitoring and forecasting weather. 
\begin{figure}[t!]
\begin{center}
\includegraphics[width=0.82\textwidth]{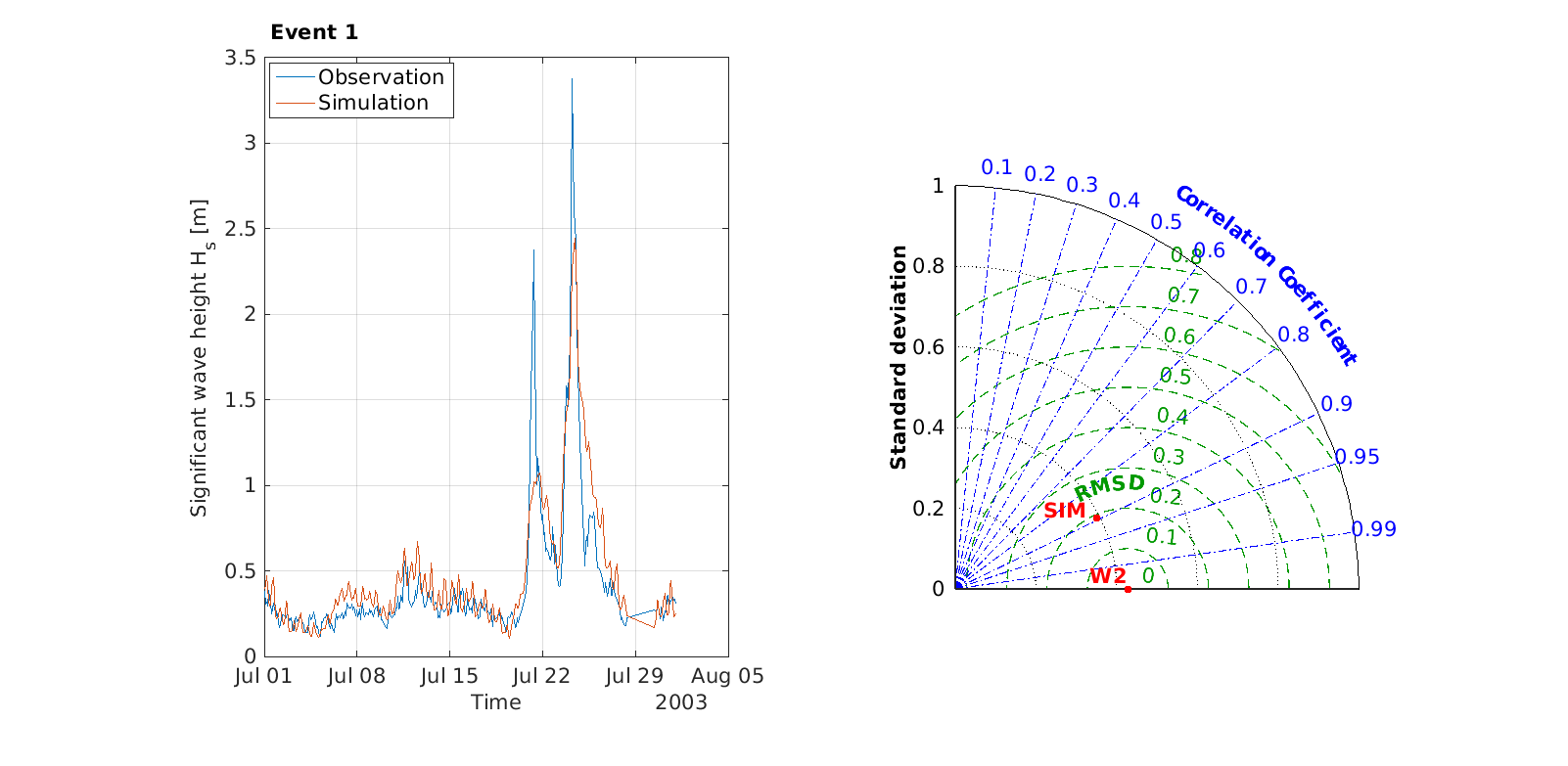}
\includegraphics[width=0.82\textwidth]{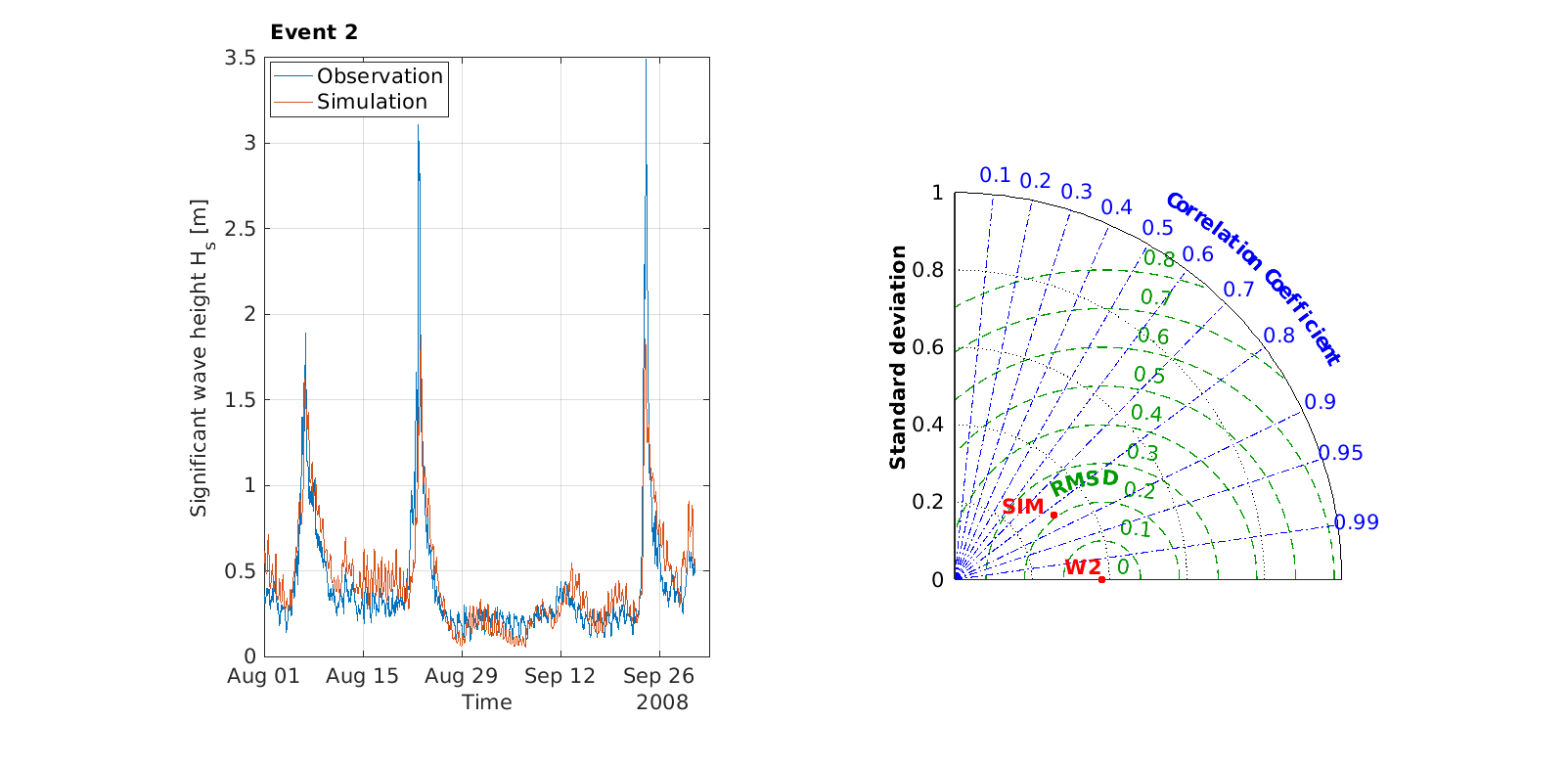}
\includegraphics[width=0.82\textwidth]{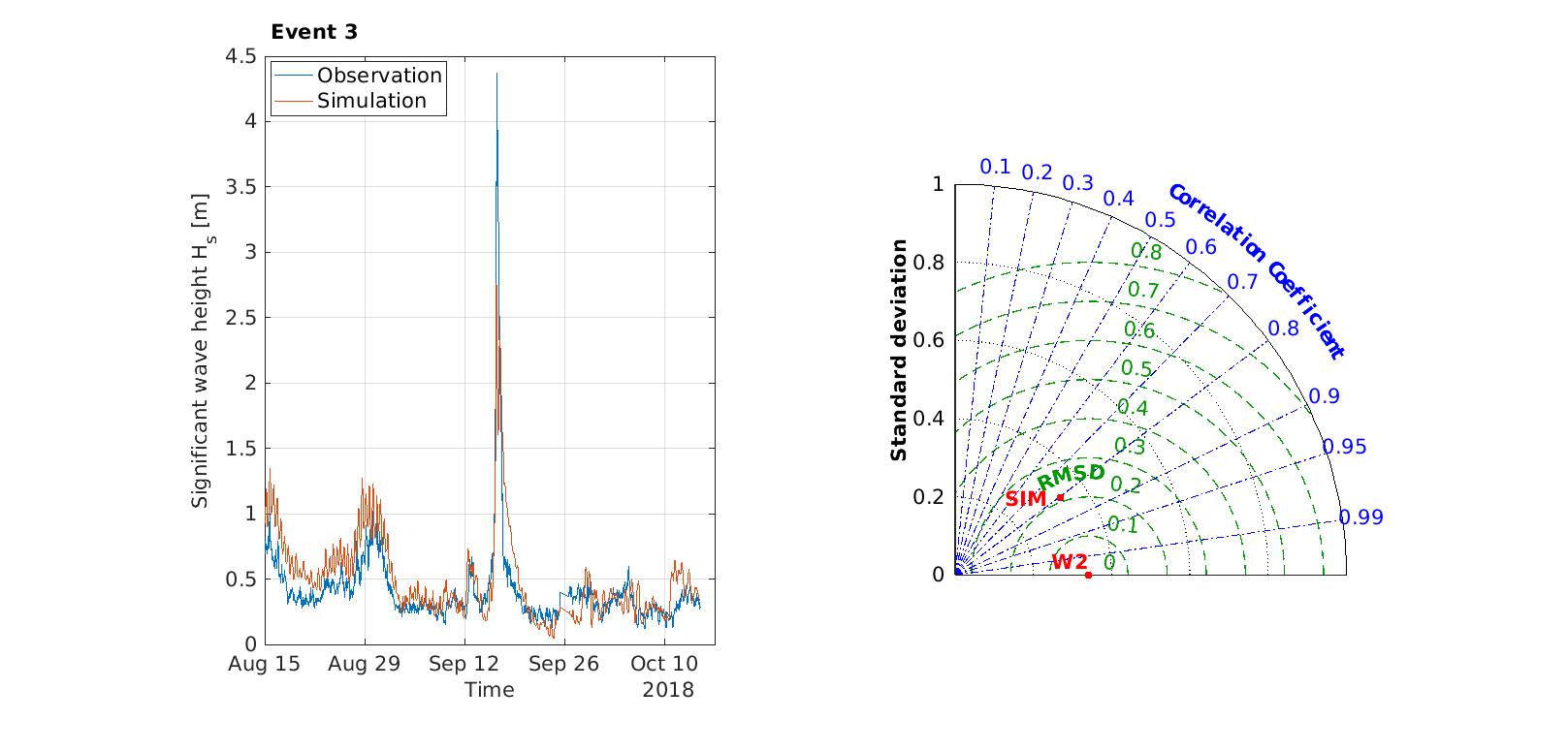}
\end{center}
\caption{Time signals of the Significant wave height plot and Taylor diagram at stations W2 for three extreme events}\label{fig:extreme}
\end{figure}
During the first period of simulation from July 1 to August 1 2003 two events were observed, close to each other. In particular, the Tropical Storm Koni (0308) developed as a tropical depression about 1000 km east-southeast of Manila on 16 July and intensified entering the South China Sea into a severe tropical storm and attained a maximum wind speed of about 100 km/h near the centre on 20 July. Koni made landfall in northern Vietnam on July 22.
Soon after Koni passed over the GBA, a new typhoon named Imbudo (0307) was recorded. Imbudo developed as a tropical depression about 730 km southwest of Guam on 17 July and intensified up to attain the grade of typhoon on July 20. The maximum wind speed was about 185 km/h near its centre on 21 July. The trajectory of Imbudo was almost parallel to the one of Koni, but northern landing near Yangjiang of western Guangdong. The maximum significant wave heights recorded by the wave buoy were about 2.4 m and 3.4 m under Koni and Imbudo, respectively, see Figure \ref{fig:extreme} panel a).

More extreme significant wave heights were recorded in the second period of simulation from August 1 to October 1 2008. During this period three events hit the SCS landing on the Chinese coastline: severe Tropical Storm Kammuri (0809) between 4 and 8 August; Typhoon Nuri (0812) between 17 and 23 August; and Typhoon Hagupit (0814) between 19 – 25 September. In particular, the two typhoons Nuri and Hagupit formed as a tropical depression over the western North Pacific and the intensified to typhoon in the SCS reaching maximum wind speeds about 180 Km/h. The results of these extreme winds led to $H_s$ between 3 and 3.5 m, see Figure \ref{fig:extreme} panel c).

The third period of simulation covered one of the most intense typhoon ever recorded by the wave bouy in the Hong Kong waters, namely the Super Typhoon Mangkhut (1822) between 7 and 17 September 2018. During the simulation period (from August 15 to October 15) other tropical storms have been reported by the Hong Kong Observatory in the annual report, namely: Severe Tropical Storm Bebinca (1816) between 9 and 17 August 2018 and Tropical Storm Barijat (1823) between 10 and 13 September 2018.
Mangkhut developed into a super typhoon on 11 September, reaching its peak intensity before making landfall over Luzon with an estimated maximum sustained wind of 250 km/h near the centre. The maximum recorded $H_s$ was about 4.4 m at WS2, see Figure \ref{fig:extreme} panel e). Due to its devastating impact, Mangkhut received a lot of attention and several studies have been published both on the analysis of the atmospheric event \citep{he2020observational,he2022observations} and on its impact on the coastal environment and urban areas \citep{yang2019comparative,he2020insights,zhou2020dynamic}.

We compared the output of the simulations in terms of significant wave height $H_s$ against the buoy measurements and evaluated the model performance using the Taylor diagram. The results of the analysis are shown in Figure \ref{fig:extreme}. Every row corresponds to a simulation period, and the left column (panels a), c) and e)) shows the time signals comparison, whereas the right column (panels b), d) and f)) shows the corresponding Taylor diagrams.
The correlation coefficient was between 0.807 and 0.895, with the lowest value for the 2018 events ($CC$ 0.807) and the highest during the 2003 events ($CC$ 0.895). The RMSD was almost invariably found to be around 0.2 (2003: RMSD 0.192 m, 2008: RMSD 0.207 m and 2018: RMSD 0.210 m), and the standard deviation was around 0.34 (2003: STD 0.393 m, 2008: STD 0.307 m and 2018: STD 0.334 m). In terms of statistical parameters, the overall performance of the numerical models could be considered fairly satisfactory. However, as expected, the maximum values of $H_s$ is not well described by the numerical simulations. Wave peaks around 2 m are reasonably captured by the model, with the only exception of the Tropical Storm Koni (July 2003). The model fails to reach wave heights of 3 m or above, which have been consistently observed during several events in the period analyzed. 

Overall, the performance with regards to the extreme events can be considered in line with previous wave hindcast \citep{perez2017gow2,shi201939}.

\subsection{Comparison with previous hindcast products and limitations}

Several global wave and storm surge hindcast are now available, see \cite{morim2022global} for a recent review and analysis of fourteen existing global wave hindcast and reanalysis products or the recent global high resolution, coupled, hindcast \citep{mentaschi2023global}. Global product are necessarily developed on relatively coarser grids from about 0.25$^{\circ}$ in specific subdomains up to 1.5$^{\circ}$ resolution \citep{hemer2013global,perez2017gow2,stopa2019sea,mentaschi2023global}.

\begin{figure}
\centerline{
 \includegraphics[width=0.5\textwidth]{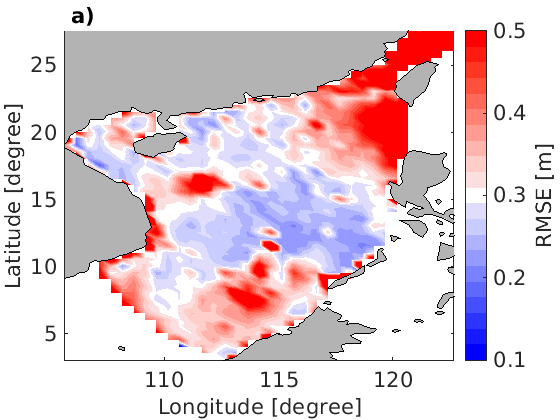}
 \includegraphics[width=0.5\textwidth]{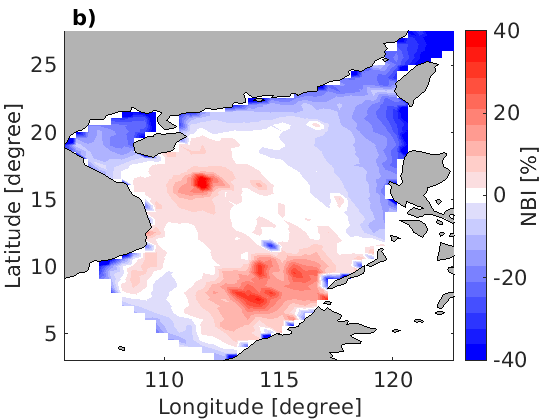}}
 \centerline{
 \includegraphics[width=0.5\textwidth]{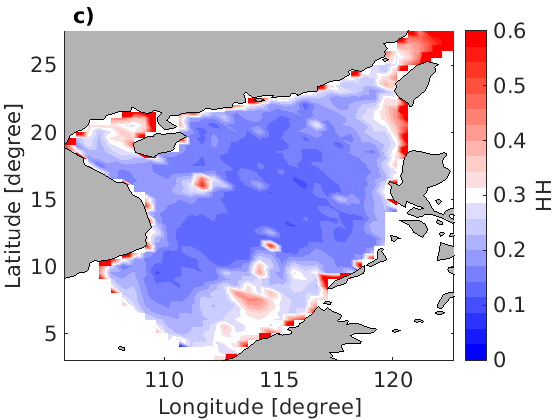}
 \includegraphics[width=0.5\textwidth]{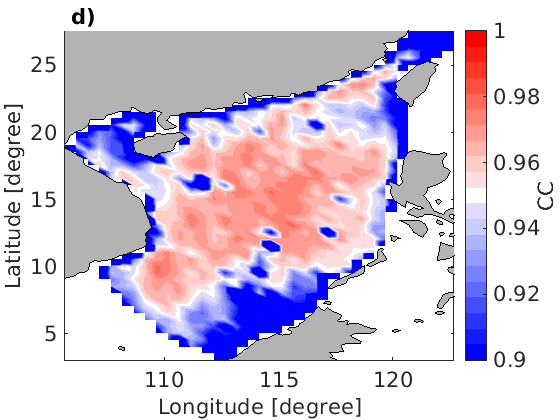}}
 
\caption{Spatial distribution of $RMSE$ (a), normalized bias $NBI$ (b), Hanna and Heinold indicator $HH$ (c) and correlation coefficient $CC$ (d) of the modeled significant wave height against satellite altimeter 
 for all tracks (1992-2019),}\label{fig:stats_maps}
\end{figure}

 Figure \ref{fig:stats_maps} shows the distribution of four statistical parameters ($RMSE$, $NBI$, $HH$ and $CC$) computed compared the simulated significant wave height and the satellite altimeter considering all tracks between 1992 and 2019, using a spatial binning of 0.5$^\circ \times$0.5$^\circ$. Comparing the present model performance in terms of the statistical parameters, the offshore skills are comparable with most of the global hindcast datasets. A detailed inspection of Figure \ref{fig:stats_maps} reveals that poorest performance are observed along the Taiwan Strait and Luzon Strait where $RMSE$, $NBI$ and $HH$ assume the highest values, whereas the correlation coefficient remains relatively high. The lower performance could be ascribed to the strong large scale currents that are not well described by the present model setup \citep{shi201939}. Other two spots of high values of $RMSE$ and $NBI$ can be observed around the Paracel Islands (around 112$^\circ$ E, 16$^\circ$ N) and the Spartly Islands (around 113$^\circ$ E, 8$^\circ$ N). The lower resolution of the mesh in open ocean does not allow for a correct representation of these two archipelagos.

Regarding the nearshore skills, as reasonable, the statistical parameters of the present study are in general better than the larger scales models \citep{perez2017gow2,mentaschi2023global}. The main reasons for the better performance of the model could be found in the higher resolution and the model configuration, namely the two ways coupling used for the present analysis. In fact, we imposed non only the atmospheric forcing, but also the tidal elevation at the open boundaries to improve the description of the non linear interactions in the nearshore zones \citep{tausia2023rapid}. 

If we consider regional hindcast of the South China Sea, a few studies have been published in the last years \citep{mirzaei2013wave,liang2016wave,shi201939}. The wave climate hindcasts were obtained using WAVEWATCHIII, SWAN and TOMAWAC with grid resolution ranging between 0.31${\circ}$-1$^{\circ}$ (offshore) to 0.15$^{\circ}$-0.01$^{\circ}$ along the coasts, with no coupling.
The wave models were validated against satellite altimeter \citep{mirzaei2013wave,shi201939} and wave buoy or ADCP measurements \citep{mirzaei2013wave,liang2016wave,shi201939}. It is worth noting that the validations against wave buoys in all three cases covers time ranges much shorter than the overall hindcast range, and the position of the wave buoys were much offshore compared the data used in the present study. The values reported in \cite{shi201939} in terms of normalized bias ($NB$) and between -0.02 and 0.03 for the 6 months in situ observations and $\pm$0.04 for the altimeter observations. The symmetrically normalized root mean square error $HH$ were reported between 0.18 and 0.23 for the in situ measurements. \cite{mirzaei2013wave} reported $RMSE$ values between 0.27 and 0.44 m with an average equal to 0.35 m for one year satellite track (2009), whereas the $RMSE$ for the four months in situ observations was 0.22 m. Finally, \cite{liang2016wave} for the period of in situ validation using seven wave buoys reported a correlation coefficient $CC$ between 0.59 and 0.93 and  $RMSE$ between 0.21 m and 0.55 m.

Comparing the present results, summarized in Tables \ref{table:all_satellite} for the mean values of all satellite tracks and \ref{table:statistics_HS_W1_W2} for the two in situ observations, we can safely state that the performance are comparable or, even, slightly better if we consider the nearshore comparison with the wave buoy. Note that the wave stations used for the present validation are located nearshore (both at water depths between 9m and 10m) and surrounded by several islands, making the comparison even more challenging. The inclusion of the nearshore wind and tidal currents in the present model, together with a fairly high resolution of the mesh, has improved the skills of the model in predicting very nearshore wave propagation.

In the future, the present dataset can be  can be improved with specific simulations aimed to better predict the cyclone-related storm surge and waves, using dedicated storm track archives for the atmospheric forcing, e.g. the IBTrACS best-track archive \citep{knapp2010international}, and the above mentioned implementations into the SCHISM‐WWMIII model package.
It is worth mentioning that the analysis conducted in this work do not account for steric effects (baroclinic conditions), which are known to improve the simulation of water levels \citep{greatbatch1994note,williams2018radiational}. As highlighted by \cite{qu2022drivers} and \cite{cheng2010steric},the contribution of steric effects in the SCS play a vital role. 
Moreover, the coupling waves and current is currently performed adopting a 2D depth-averaged approach for the circulation model. A three dimensional model would improve the representation of the complicated open ocean and coastal circulation, most probably leading to a better prediction in terms of wave climate.

\section{Conclusions}
In this study the fully coupled SCHISM-WWMIII model was considered in order to generate a comprehensive 53-year wave hindcast spanning from 1970 to 2022 in the majority of the South China Sea (SCS). The detailed unstructured mesh and the high resolution in the Hong Kong waters allowed the model to successfully capture the non-linear processes caused by the mutual effects of waves and currents, particularly in shallow regions.
Five different meshes were considered in the sensitivity analysis and the most performing mesh, which minimized computational time while optimizing numerical accuracy was utilized for the numerical simulations. The comparison between model outputs ( $H_s$, $T_p$, $D_m$ and $h$) and observations showed close conformity offshore and slightly reduced accuracy in nearshore areas, known to be notoriously challenging because of the complexity induced by the coastline and the bathymetry.
In general, the statistical indicators related to water level and significant wave height demonstrated high values of correlation and relatively low errors. However, peak wave periods and mean wave directions appeared to be more challenging. 
The observation data from wave stations and satellites, as well as wind velocities and water level forcing, were retrieved from online databases.

Further research effort could focus on the analysis of the coastal processes amplified by the presence of small islands and complex coastlines. Employing a more detailed mesh and bathymetry could also lead to improved statistical indicators and enhance the performance of the model. This study serves as a preliminary stage upon which more solid studies about wave climate, wave trend and wave energy assessment can be carried out.

\bibliographystyle{unsrtnat}






\end{document}